\documentclass[aps,pre,twocolumn,showpacs]{revtex4}
\usepackage{graphicx}
\usepackage{amsmath,amscd,amssymb}
\usepackage{color}

\newcommand{\be}{\begin{equation}}
\newcommand{\ee}{\end{equation}}
\newcommand{\bea}{\begin{eqnarray}}
\newcommand{\eea}{\end{eqnarray}}
\newcommand{\lan}{\left\langle}
\newcommand{\ran}{\right\rangle}
\newcommand{\br}{\mathbf{r}}

\newcommand{\bk}{\mathbf{k}}
\newcommand{\e}{\varepsilon}

\newcommand{\tv}{\tilde{v}}

\newcommand{\bz}{\bar{z}}
\newcommand{\bd}{\bar{\Delta}}

\begin{document}

\title{Electrostatic correlations in inhomogeneous charged fluids beyond loop expansion}

\author{Sahin Buyukdagli$^{1}$\footnote{email:~\texttt{sahin\_buyukdagli@yahoo.fr}}, C. V. Achim$^{1}$\footnote{email:~\texttt{cristian.v.achim@gmail.com}}, and T. Ala-Nissila$^{1,2}$\footnote{email:~\texttt{Tapio.Ala-Nissila@aalto.fi}}}
\affiliation{$^{1}$Department of Applied Physics and COMP center of Excellence, Aalto University School of Science, P.O. Box 11000, FI-00076 Aalto, Espoo, Finland\\
$^{2}$Department of Physics, Brown University, Providence, Box 1843, RI 02912-1843, U.S.A.}
\date{\today}

\begin{abstract}
Electrostatic correlation effects in inhomogeneous symmetric electrolytes are investigated within a previously developed electrostatic self-consistent (SC) theory (R.R. Netz and H. Orland, Eur. Phys. J. E \textbf{11}, 301 (2003)). To this aim, we introduce two computational approaches that allow to solve the SC equations beyond the loop expansion. The first method is based on a perturbative Green's function technique, and the second one is an extension of a previously introduced semiclassical approximation for single dielectric interfaces to the case of slit nanopores. Both approaches can handle the case of dielectrically discontinuous boundaries where the one-loop theory is known to fail. By comparing the theoretical results obtained from these schemes with the results of the MC simulations that we ran for ions at neutral single dielectric interfaces, we first show that the weak coupling (WC) Debye-Huckel (DH) theory remains quantitatively accurate up to the bulk ion density $\rho_b\simeq0.01$ M, whereas the SC theory exhibits a good quantitative accuracy up to $\rho_b\simeq0.2$ M, thus improving the accuracy of the DH theory by one order of magnitude in ionic strength. Furthermore, we compare the predictions of the SC theory with previous MC simulation data for charged dielectric interfaces and show that the proposed approaches can also accurately handle the correlation effects induced by the surface charge in a parameter regime where the mean-field (MF) result significantly deviates from the MC data. Then, we derive from the perturbative SC scheme the one-loop theory of asymmetrically partitioned salt systems around a dielectrically homogeneous charged surface. It is shown that correlation effects originate in these systems from a competition between the salt screening loss at the interface driving the ions to the bulk region, and the interfacial counterion screening excess attracting them towards the surface. This competition can be quantified in terms of the characteristic surface charge $\sigma_s^*=\sqrt{2\rho_b/(\pi\ell_B)}$, where $\ell_B=7$ {\AA} is the Bjerrum length. In the case of weak surface charges $\sigma_s\ll\sigma_s^*$ where counterions form a diffuse layer, the interfacial salt screening loss is the dominant effect. As a result, correlation effects decrease the MF density of both coions and counterions. With an increase of the surface charge towards $\sigma_s^*$, the surface-attractive counterion screening excess starts to dominate, and correlation effects amplify in this regime the MF density of both type of ions. However, in the regime $\sigma_s>\sigma_s^*$, the same counterion screening excess also results in a significant decrease of the electrostatic MF potential. This reduces in turn the MF counterion density far from the charged surface. We also show that for $\sigma_s\gg\sigma_s^*$, electrostatic correlations result in a charge inversion effect. However, the electrostatic coupling regime where this phenomenon takes place should be verified with MC simulations since this parameter regime is located beyond the validity range of the one-loop theory.
\end{abstract}

\maketitle

\section{Introduction}

The Poisson-Boltzmann (PB) formalism developed a century ago by Gouy~\cite{Gouy} and Chapman~\cite{Chapman} is still considered today as the elemental theoretical description of electrostatic effects in various microscopic systems. The solution of the PB equations for charged macromolecules immersed in salt solutions allows for example to  determine protein folding pathways~\cite{mol} or to understand the stability of DNA-binding proteins during their diffusion along DNA molecules~\cite{DNA}. One can also mention the microfluidic devices where the solute velocity is derived from the coupled solution of the PB and Navier-Stokes equations~\cite{micro}. Being a MF theory, the PB formalism neglects however electrostatic correlation associated with the interaction of the charged fluid with the system boundaries. It is thus clear that the PB equation is a crude approximation for dielectrically discontinuous systems such as water-air or water-membrane interfaces where the electrolyte-surface interactions can significantly exceed the thermal energy $k_BT$ in the proximity of the interface.

The pioneering consideration of electrostatic correlation effects in inhomogeneous charged systems is doubtlessly Wagner's interpretation of the surface tension excess of water with added salt in terms of the screened image charge interactions~\cite{wagner}. This theoretical framework that allowed Onsager and Samaras to derive their celebrated limiting law~\cite{onsager} was later improved in Ref.~\cite{japan} by accounting for the non-uniform shielding of image interactions. Within a Wentzel-Kramers-Brillouin (WKB) approximation, the author ingeniously evaluated the modification of the ionic self-energy and the surface tension by the interfacial variations of the ionic screening, which improved the agreement with experimental surface tension data.

Correlation effects induced by the polarization charges at dielectric interfaces are also relevant to various industrial applications, among which one can mention water purification and desalination processes in artificial nanofiltration technology. Non-linear electrostatic SC equations for confined electrolytes were derived in Ref.~\cite{SCrussian}  within the Debye closure of the Bogoliubov-Born-Green-Kirkwood-Yvon (BBGKY) hierarchical equations. Approximative solutions of these SC equations describe the selectivity of nanofiltration membranes in terms of a cooperation between the dielectric exclusion mechanism induced by image forces and the Donnan rejection driven by the membrane surface charge~\cite{yarosch,YarII}. They are frequently used today in order to predict experimental salt rejection rates in artificial nanofiltration processes~\cite{Szymczyk}.

The validity regime of the mathematical framework of these theories remained however unclear for several decades. The field theoretic formulation of the heterogeneous Coulomb fluid model indeed provides some control over the approximations involved in the consideration of correlation effects in nanoscale systems. The field has witnessed a dramatic growth during the last two decades. To give a non-exhaustive list, one can mention for example the consideration of  electrostatic correlations in macromolecular forces for counterion liquids~\cite{PodWKB} and symmetric electrolytes~\cite{attard} at the gaussian level. The one-loop corrections to the density of counterions in contact with charged walls was also introduced in Ref.~\cite{netzcoun}.

Since the electrostatic coupling parameter increases with the ion density as $\Gamma\propto\sqrt{\rho_b}$, the exploration of the parameter regime beyond the dilute electrolyte limit necessitates the consideration of non-linear effects neglected in gaussian field theories. As we will explicitly show in the present work, this is the regime where SC theories become relevant. The SC equations of Ref.~\cite{SCrussian} were rederived in Ref.~\cite{netzvar} within the field theoretic formulation of symmetric electrolytes. As stressed by the authors, these coupled SC equations are too complicated to be solved even numerically, and one has to make use of additional approximations in order to explore the underlying physics. These SC equations were solved in Ref.~\cite{Lee} within a WKB-like approach in order to investigate ionic partitions around charged dielectric cylinders. However, we note that the approach used therein is not exactly a WKB method since the author did not make use of the WKB ansatz as in Ref.~\cite{japan}, but rather solved the SC equations by assuming that the local screening parameter in these equations does not vary with the spatial coordinate. A restricted variational theory was also proposed in Refs.~\cite{hatlo} in order to understand non-linear effects in the process of dielectric exclusion from neutral slit nanopores. Furthermore, it was shown in Ref.~\cite{hatlorev} that the consideration of the interfacial ionic screening deficiency in this variational theory can considerably improve the agreement with MC simulations. An efficient restricted variational approach able to handle correlation effects induced by the surface charge from weak to strong coupling regime was also proposed in Ref.~\cite{hatloepl}.

We introduced in Ref.~\cite{PRE} a simpler variational approach for ions confined in charged slit nanopores that was able to handle the membrane charge with a good agreement with MC simulation results beyond the MF limit. We also applied this approach to cylindrical ion channels in order to show that the complications resulting from the curvature of the dielectric interface can explain the ionic current fluctuations in biological and artificial nanopores~\cite{PRL}. We then extended the method by taking into account the excluded volume of ions in order to study excluded volume effects in the dielectric exclusion mechanism~\cite{jstat} and macromolecular interactions~\cite{JCP2}. Using similar ideas, we finally derived in Ref.~\cite{epl} a non-mean-field dipolar PB equation in order to show that the interfacial solvent depletion at low dielectric surfaces can solely explain the low values of the experimental differential capacitance data of carbon based materials.

The weakness of the restricted variational approaches in Refs.~\cite{hatlo,hatlorev,hatloepl,PRE,PRL,jstat,JCP2,epl} is that the restricted variational ansatz determines the nature of the final solution. The main goal of the present work is to overcome this limitation by solving the general SC equations of Ref.~\cite{netzvar} within two new computational approaches beyond the loop expansion. The latter point will be shown to be crucial for understanding electrostatic correlation effects in dielectrically inhomogeneous systems where the one-loop theory is known to fail~\cite{David}.

The article is organized as follows. We present in Section~\ref{redSC} an alternative derivation of the SC equations of Ref.~\cite{netzvar}. We then introduce in Section~\ref{yuh} two approximative methods to solve these equations. The first method is based on an expansion of the formal inversion of the SC equations in powers of the fluctuating particle and charge density excesses around the weak-coupling theory. This approach is formally equivalent to the iterative solution of Hartree equations in condensed matter physics~\cite{Alexei}. The second method is an extension of the previously introduced WKB solution of these equations~\cite{japan} at simple dielectric interfaces to the case of slit nanopores.  We first compare in Section~\ref{res} the predictions of these schemes for ion densities with the results of MC simulations that we ran for ions at neutral dielectric interfaces in order to establish the validity domain of the WC and SC theories. Then, we test the validity of previous variational schemes for ions in slit nanopores, and also investigate within the WKB approach the interaction between a charged rigid polymer and a dielectric wall. Furthermore, by comparisons with previous MC simulation data for ions in contact with a charged surface, we show that SC equations can handle correlation effects at charged dielectric interfaces beyond the WC parameter regime. Finally, we derive from the SC scheme the one-loop theory of asymmetrically distributed salt solutions that we thoroughly examine. This one-loop calculation bridges a gap between the DH theory of symmetric electrolytes at neutral interfaces~\cite{netzdh} and the one-loop theory of counterion liquids in contact with charged interfaces~\cite{netzcoun}. Possible extensions of the concepts introduced in this article are discussed in the Conclusion part.

\section{Rederivation of SC equations for symmetric electrolytes}
\label{redSC}

We will derive in this part the self-consistent equations of Ref.~\cite{netzvar} in a shortcut way that does not require the evaluation of the Grand potential of the inhomogeneous electrolyte system. The grand canonical partition function of a symmetric electrolyte composed of two species of valency $\pm q$ with $q>0$, and fugacity $\Lambda_i$ is given by a functional integral over a fluctuating electrostatic potential $\phi(\br)$, $Z_G=\int \mathcal{D}\phi\;e^{-H[\phi]}$, with the Hamiltonian functional~\cite{netzvar}
\bea\label{HamFunc}
H[\phi]&=&\int\mathrm{d}\br\left[\frac{\left[\nabla\phi(\br)\right]^2}{8\pi\ell_B(\br)}-i\sigma(\br)\phi(\br)\right]\\
&&-2\Lambda_i \int\mathrm{d}\br e^{E_i-V_w(\br)}\cos\left[q\phi(\br)\right],\nonumber
\eea
where $\sigma(\br)$ is a fixed charge distribution, $\ell_B(\br)=e^2/(4\pi k_BT\e(\br))$ the Bjerrum length, $\e(\br)$ the static dielectric permittivity profile of the medium, and $e$ the elementary charge. The wall potential $V_w(\br)$ restricts the space volume accessible to ions. Furthermore, $E_i=\frac{q^2}{2}v_c^b(\br-\br')|_{\br=\br'}$ is the self energy of ions in salt-free water, and $v_c^b(r)=\ell_B/r$  the Coulomb potential in a bulk solvent, with $\ell_B=7$ {\AA} the Bjerrum length in a bulk electrolyte at ambient temperature $T=300$ K. We finally note that in the present work, all energies are expressed in units of the thermal energy $k_BT$, the surface charge in units of the elementary charge $e$, and the dielectric permittivities in units of the dielectric permittivity of the air $\e_0$.

Our starting point is the compact form of the Schwinger-Dyson equation
\be\label{eq1}
\int \mathcal{D}\phi\;\frac{\delta}{\delta\phi(\br)}e^{-H[\phi]+\int\mathrm{d}\br J(\br)\phi(\br)}=0,
\ee
where $J(\br)$ is an external current. A rigorous proof of the equality~(\ref{eq1}) can be found for example in Ref.~\cite{justin}. We will derive from Eq.~(\ref{eq1}) two Ward identities relating the external electrostatic potential and the electrostatic propagator to higher order correlation functions. By acting now on the exponential with the functional derivative and setting $J(\br)=0$, one gets the following equation for the electrostatic potential,
\be\label{eq2}
\frac{k_BT}{e^2}\nabla\e(\br)\nabla\lan\phi(\br)\ran+i\sigma(\br)-2\Lambda_iq e^{E_i-V_w(\br)}\lan\sin\left[q\phi(\br)\right]\ran=0.
\ee
We note that this equation was obtained in Ref.~\cite{1loop}. Furthermore, taking the functional derivative of Eq.~(\ref{eq1}) with respect to $J(\br')$ and setting again $J(\br)=0$, we obtain a new relation
\bea\label{eq3}
&&\frac{k_BT}{e^2}\nabla\e(\br)\nabla\lan\phi(\br)\phi(\br')\ran+i\sigma(\br)\lan\phi(\br')\ran\\
&&-2\Lambda_iq e^{E_i-V_w(\br)}\lan\phi(\br')\sin\left[q\phi(\br)\right]\ran=-\delta(\br-\br').\nonumber
\eea
The approximation now consists in evaluating the averages over the fluctuations in Eqs.~(\ref{eq2}) and~(\ref{eq3}) with the effective Hamiltonian of the most general quadratic dependence on the fluctuating potential $\phi(\br)$ instead of the non-linear one in Eq.~(\ref{HamFunc}),
\be\label{eq4}
H_0=\frac{1}{2}\int_{\br,\br'}\left[\phi(\br)-i\phi_0(\br)/q\right]
v^{-1}_0(\br,\br')\left[\phi(\br')-i\phi_0(\br')/q\right],
\ee
where the external electrostatic potential $\phi_0(\br)\equiv-iq\lan\phi(\br)\ran$ and the inverse of the kernel $v(\br,\br')\equiv\lan\phi(\br)\phi(\br')\ran-\lan\phi(\br)\ran\lan\phi(\br')\ran$ are solutions of Eqs.~(\ref{eq2}) and~(\ref{eq3}). The evaluation of the statistical averages in  Eqs.~(\ref{eq2}) and~(\ref{eq3}) with the effective Hamiltonian Eq.~(\ref{eq4}) finally yields the self-consistent equations of Ref.~\cite{netzvar},
\bea\label{eq5}
&&\nabla\e(\br)\nabla\phi_0(\br)-\e(\br)\kappa_b^2e^{-V_w(\br)-\frac{q^2}{2}\delta v(\br,\br)}\sinh\left[\phi_0(\br)\right]\\
&&=-\frac{e^2q}{k_BT}\sigma(\br)\nonumber\\
\label{eq6}
&&\nabla\e(\br)\nabla v(\br,\br')-\e(\br)\kappa_b^2e^{-V_w(\br )-\frac{q^2}{2}\delta v(\br,\br)}\cosh\left[\phi_0(\br)\right]v(\br,\br')\nonumber\\
&&=-\frac{e^2}{k_BT}\delta(\br-\br'),
\eea
where we took into account the relation between the ion fugacity and the bulk density $\Lambda_i=\rho_{bi}e^{-\frac{q^2}{2}\kappa_b\ell_B}$~\cite{hatlo,PRE}. We also defined the ionic self-energy dressed with electrostatic correlations in the form
\be\label{eq7}
\delta v(\br,\br)=\ell_B\kappa_b+ v(\br,\br)-v_c^b(0),
\ee
where we introduced the bulk screening parameter as $\kappa_b^2=8\pi q^2\ell_B\rho_b$. We also note that the local ion densities are given by~\cite{PRE}
\be\label{den}
\rho_\pm(\br)=\rho_{bi}e^{-V_w(\br)-\frac{q^2}{2}\delta v(\br,\br)\mp\phi_0(\br)}.
\ee

The relation Eq.~(\ref{eq5}) is a modified PB equation for the fluctuating external potential induced by the fixed surface charge around the MF potential. The second differential equation~(\ref{eq6}) is a generalized Laplace equation that accounts for the local screening of the electrostatic propagator by mobile ions. In the next section, we will develop two approximative methods to solve these equations.

\section{Computational schemes}
\label{yuh}

We introduce in this section two computational schemes for solving the SC equations~(\ref{eq5}) and~(\ref{eq6}). The first scheme consists in solving these equations by expanding their formal inversion around the one-loop electrostatic Green's function and the non-linear MF potential in fluctuating excess charge and particle densities.  The second scheme based on a WKB approach is an extension of a previously introduced solution of the equation~(\ref{eq6}) for single neutral interfaces~\cite{japan} to the more complicated case of neutral slit nanopores. The results of the theoretical approaches introduced in this section will be compared in the next section with extensive MC simulation data in order to determine the validity domain of Eqs.~(\ref{eq5}) and~(\ref{eq6}).

\subsection{Perturbative solution of SC equations}
\label{anan}

We present in this section an iterative solution of the SC equations~(\ref{eq5}) and~(\ref{eq6}) around the MF external potential $\varphi(r)$ and the one-loop Green's function $v_0(\br,\br')$, respectively solutions of the equations~\cite{1loop}
\bea\label{eq333}
&&\nabla\e(\br)\nabla\varphi(\br)-\e(\br)\kappa_b^2e^{-V_w(\br)}\sinh\left[\varphi(\br)\right]=-\frac{e^2q}{k_BT}\sigma(\br)\nonumber\\
&&\\
\label{eq334}
&&\left\{\nabla\e(\br)\nabla-\e(\br)\kappa_b^2e^{-V_w(\br)}\cosh\left[\varphi(\br)\right]\right\}v_0(\br,\br')\\
&&=-\frac{e^2}{k_BT}\delta(\br-\br').\nonumber
\eea
The first step consists in injecting into Eq.~(\ref{eq5}) the fluctuating part of the external potential $\psi(\br)\equiv\phi_0(\br)-\varphi(\br)$, and rearranging the resulting equation for $\psi(\br)$ with Eq.~(\ref{eq6}) in the form
\bea\label{eq111}
&&\left\{\nabla\e(\br)\nabla-\e(\br)\kappa_b^2e^{-V_w(\br)}\cosh\left[\varphi(\br)\right]\right\}\psi(\br)\\
&&=\e(\br)\kappa_b^2\delta\sigma(\br)\nonumber\\
\label{eq112}
&&\left\{\nabla\e(\br)\nabla-\e(\br)\kappa_b^2e^{-V_w(\br)}\cosh\left[\varphi(\br)\right]\right\}v(\br,\br')\\
&&=-\frac{e^2}{k_BT}\delta(\br-\br')+\e(\br)\kappa_b^2\delta n(\br)v(\br,\br'),\nonumber
\eea
where we introduced respectively the fluctuating charge and particle density excesses as
\bea\label{eq335}
\delta\sigma(\br)&=&\left\{n(\br)\sinh\left[\phi_0(\br)\right]-\sinh\left[\varphi(\br)\right]\right.\\
\label{eq336}
&&\left.-\cosh\left[\varphi(\br)\right]\psi(\br)\right\}e^{-V_w(\br)}\nonumber\\
\delta n(\br)&=&\left\{n(\br)\cosh\left[\phi_0(\br)\right]-\cosh\left[\varphi(\br)\right]\right\}e^{-V_w(\br)},\nonumber\\
\eea
with the Boltzmann factor $n(\br)=e^{-\frac{q^2}{2}\delta v(\br,\br)}$. Using now Eq.~(\ref{eq334}), the relations~(\ref{eq111}) and~(\ref{eq112}) can be inverted as
\bea\label{eq331}
\psi(\br)&=&-2\rho_bq^2\lambda_{\phi}\int\mathrm{d}\br_1v_0(\br,\br_1)\delta\sigma(\br_1)\\
\label{eq332}
v(\br,\br')&=&v_0(\br,\br')\\
&&-2\rho_bq^2\lambda_v\int\mathrm{d}\br_1v_0(\br,\br_1)\delta n(\br_1)v(\br_1,\br'),\nonumber
\eea
where we introduced the expansion parameters $\lambda_{\phi}$ and $\lambda_v$ that will enable us to keep track of the perturbative order. These parameters will be set to unit at the end of the perturbative expansion.

For charged liquids bounded by charged planar interfaces located within the $(x,y)$ plan, the translational symmetry considerably simplifies the problem. In this geometry, the external potential becomes simply a function of the coordinate $z$, and the electrostatic Green's function can be expanded in 2D Fourier basis as
\be\label{fourker}
v(\br,\br')=\int\frac{d^2\bk}{4\pi^2}e^{i\bk\cdot\br_{\|}}\tv(z,z',k).
\ee
We now expand the Green's function and the fluctuating external potential in $\lambda_v$ and $\lambda_{\phi}$,
\bea\label{eq337}
&&v(\br,\br')=v_0(\br,\br')+\sum_{n,m>0}\lambda_v^n\lambda_{\phi}^m\delta v_{nm}(\br,\br')\\
\label{eq338}
&&\psi(z)=\sum_{n,m>0}\lambda_v^n\lambda_{\phi}^m\psi_{nm}(z).
\eea
Injecting these expansions into Eqs.~(\ref{eq331})-(\ref{eq332}) and Eqs.~(\ref{eq337})-(\ref{eq338}), and keeping only the terms up to the order $\lambda_v^2$, $\lambda_{\phi}^2$, and $\lambda_v\lambda_{\phi}$, we get $\delta v_{01}(\br,\br')=\delta v_{02}(\br,\br')=0$, $\psi_{10}(z)=\psi_{20}(z)=0$, and
\bea\label{eq339}
\delta v_{10}(\br,\br')&=&-2\rho_bq^2\int\mathrm{d}z_1\delta n_0(z_1)I_2(\br,\br',z_1)\\
\label{eq340}
\delta v_{11}(\br,\br')&=&-2\rho_bq^2\int\mathrm{d}z_1 n_0(z_1)\sinh\left[\varphi(z_1)\right]\psi_{01}(z_1)\nonumber\\
&&\hspace{1.5cm}\times I_2(\br,\br',z_1)\\
\label{eq341}
\delta v_{20}(\br,\br')&=&\rho_bq^4\int\mathrm{d}z_1 n_0(z_1)\cosh\left[\varphi(z_1)\right]\delta v_{10}(z_1)\nonumber\\
&&\hspace{1.5cm}\times I_2(\br,\br',z_1)\\
&&+\left(2\rho_bq^2\right)^2\int\mathrm{d}z_1\mathrm{d}z_2\delta n_0(z_1)\delta n_0(z_2)\nonumber\\
&&\hspace{1.5cm}\times I_3(\br,\br',z_1,z_2)
\eea
for the corrections to the Green's function, and
\bea
\label{eq342}
&&\psi_{01}(z)=-2\rho_bq^2\int\mathrm{d}z_1\tv_0(z,z_1,0)\delta \sigma_0(z_1)\\
\label{eq343}
&&\psi_{11}(z)=\rho_bq^4\int\mathrm{d}z_1\tv_0(z,z_1,0) n_0(z_1)\sinh\left[\varphi(z_1)\right]\nonumber\\
&&\hspace{3cm}\times\delta v_{10}(z_1)\\
\label{eq344}
&&\psi_{02}(z)=-2\rho_bq^2\int\mathrm{d}z_1\tv_0(z,z_1,0)\delta n_0(z_1)\psi_{01}(z_1)\nonumber\\
\eea
for the fluctuating potential. We note that we used above the notation $\delta v_{nm}(z)=\delta v_{nm}(\br,\br)$ for the equal point excess Green's function, and introduced the auxiliary functions
\bea
&&I_2(\br,\br',z_1)=\int\frac{\mathrm{d}^2\bk}{4\pi^2}e^{i\bk\cdot\left(\br_\parallel-\br'_\parallel\right)}\tv_0(z,z_1,k)\tv_0(z_1,z',k)\nonumber\\
&&\\
&&I_3(\br,\br',z_1,z_2)=\int\frac{\mathrm{d}^2\bk}{4\pi^2}e^{i\bk\cdot\left(\br_\parallel-\br'_\parallel\right)}\tv_0(z,z_1,k)\tv_0(z_1,z_2,k)\nonumber\\
&&\hspace{4cm}\times\tv_0(z_2,z',k),
\eea
and
\bea\label{sigper}
&&\delta\sigma_0(z)=\left[n_0(z)-1\right]\sinh\left[\varphi(z)\right]e^{-V_w(z)}\\
\label{nper}
&&\delta n_0(z)=\left[n_0(z)-1\right]\cosh\left[\varphi(z)\right]e^{-V_w(z)},
\eea
with $n_0(z)=e^{-\frac{q^2}{2}\delta v_0(z)}$. Finally, the expansion of the ion densities in Eq.~(\ref{den}) yields at the same perturbative level the following expression,
\bea\label{denper}
\rho_\pm(z)&=&\rho_\pm^{(0)}(z)\left\{1-\lambda_v\frac{q^2}{2}\delta v_{10}(z)\mp\lambda_{\phi}\psi_{01}(z)\right.\\
&&+\lambda_v\lambda_{\phi}\left[-\frac{q^2}{2}\delta v_{11}(z)\mp\psi_{11}(z)\pm\frac{q^2}{2}\delta v_{10}(z)\psi_{01}(z)\right]\nonumber\\
&&+\lambda_v^2q^2\left[\frac{1}{8}\delta v_{10}^2(z)-\frac{1}{2}\delta v_{20}(z)\right]\nonumber\\
&&\left.+\lambda_{\phi}^2\left[\frac{1}{2}\psi_{01}^2(z)\mp\psi_{02}(z)\right]\right\},\nonumber
\eea
where the zeroth order ion density is given by
\be\label{den0}
\rho_\pm^{(0)}(z)=\rho_{bi}e^{-V_w(z)-\frac{q^2}{2}\delta v_0(z)\mp\varphi(z)}.
\ee
We note that the above equations can be easily generalized to the case of asymmetrical electrolytes. We will derive below the one-loop propagator needed to compute the correction terms in Eqs.~(\ref{eq339})-(\ref{eq344}) for the electrostatic Green's function and the external potential.

\subsubsection{Neutral interfaces}
\label{baban}

In the case of neutral interfaces with $\sigma(z)=0$, the external potential and the corrective cross term of the Green's function both vanish, that is $\phi_0(z)=0$ and $\delta v_{11}=0$. Furthermore, we note that for electrolytes confined between two planar interfaces located at $z=0$ and $z=d$, and separating the solvent part with dielectric permittivity $\e_w=80$ from the membrane matrix with permittivity $\e_m<\e_w$, the wall potential is defined as $V_w(z)=0$ if $0\leq z\leq d$, and $V_w(z)=\infty$ if $z<0$ or $z>d$.

For the same slit geometry, the Fourier transform of the DH Green's function required to compute the correction terms in Eqs.~(\ref{eq339})-(\ref{eq341}) reads~\cite{netzvdw}
\bea\label{eq01}
\tilde v_0(z,z',k)&=&\frac{2\pi\ell_B}{p_b}\left\{e^{-p_b|z-z'|}\right.\\
&&+\frac{\Delta_b}{1-\Delta_b^2e^{-2p_bd}}\left[e^{-p_b(z+z')}+e^{p_b(z+z'-2d)}\right.\nonumber\\
&&\hspace{2cm}\left.\left.+2\Delta_b e^{-2p_bd}\cosh\left(p_b|z-z'|\right)\right]\right\},\nonumber
\eea
where we have defined
\bea\label{pb}
p_b&=&\sqrt{\kappa_b^2+k^2}\\
\label{deltab}
\Delta_b&=&\frac{\e_wp_b-\e_mk}{\e_wp_b+\e_mk}.
\eea
We also note that for ions confined in the nanoslit, the spatial integrals in Eqs.~(\ref{eq339})-(\ref{eq341}) run over the interval $0\leq z\leq d$. Moreover, for a single neutral interface obtained from the slit pore geometry in the limit $d\to\infty$, the weak-coupling Green's function~(\ref{eq01}) reduce to
\be\label{eq02}
\tilde v_0(z,z',k)=\frac{2\pi\ell_B}{p_b}\left\{e^{-p_b|z-z'|}+\Delta_be^{-p_b(z+z')}\right\},
\ee
and the integrals in Eqs.~(\ref{eq339})-(\ref{eq341}) have to be carried out over the right half space $z\geq0$ occupied by the electrolyte.

\subsubsection{Charged single interface}
\label{ninen}

We will derive in this part the zeroth order Green's function $\tv_0(z,z',k)$ and the associated ionic self energy $\delta v_0(z)$ needed to compute the corrective terms in Eqs.~(\ref{eq339})-(\ref{eq344}) for the case of a dielectric interface located at $z=0$, and carrying a negative surface charge $\sigma(z)=-\sigma_s\delta(z)$ with $\sigma_s>0$. For this geometry, the MF potential solution of Eq.~(\ref{eq333}) is given by~\cite{Isr}
\be\label{pmf}
\varphi(z)=2\ln\left[\frac{1-e^{-\kappa_b(z+z_0)}}{1+e^{-\kappa_b(z+z_0)}}\right],
\ee
where we used the same notation as in Ref.~\cite{1loop} for the characteristic length of the interfacial counterion layer $z_0=-\ln(\gamma_c(s))/\kappa_b$ , with the auxiliary parameter $\gamma_c(s)=\sqrt{s^2+1}-s$, the dimensioneless parameter $s=\kappa_b\mu$, and the Gouy-Chapman length $\mu=1/(2\pi q\ell_B\sigma_s)$.

By injecting first the Fourier transform of the Green's function Eq.~(\ref{fourker}) into the Green's equation~(\ref{eq334}), we obtain
\bea\label{eq01}
&&\frac{\partial}{\partial z}\e(z)\frac{\partial}{\partial z}\tv_0(z,z',k)\\
&&-\e(z)\theta(z)\left\{p_b^2+2\kappa_b^2\mathrm{csch}^2\left[\kappa_b(z+z_0)\right]\right\}\tv_0(z,z',k)\nonumber\\
&&=-\frac{e^2}{k_BT}\delta(z-z').
\eea
For the ion source located in the water medium $z'\geq0$, the solution of this equation reads $\tv(z,z',k)=c_1e^{kz}\theta(-z)+\left[c_2h_+(z)+c_3h_-(z)\right]\theta(z)\theta(z'-z)$, where the homogeneous solutions for $z>0$ were found  in Ref.~\cite{1loop} in the form
\be\label{solpar}
h_\pm(z)=e^{\pm p_bz}\left\{1\mp\frac{\kappa_b}{p_b}\coth\left[\kappa_b(z+z_0)\right]\right\},
\ee
The coefficients $c_i$ for our system of impenetrable dielectric wall have to be computed by imposing the usual boundary conditions associated with the continuity of the electrostatic potential and the displacement field~\cite{netzvdw},
\bea\label{boun1}
&&\tv_0\left(z=\Sigma_-\right)=\tv\left(z=\Sigma_+\right)\\
\label{boun2}
&&\e(z)\left.\frac{\partial\tv_0}{\partial z}\right|_{z=\Sigma_-}=\e(z)\left.\frac{\partial \tv_0} {\partial z}\right|_{z=\Sigma_+}\\
\label{boun3}
&&\left.\frac{\partial\tv_0}{\partial z}\right|_{z=z'_+}-\left.\frac{ \partial \tv_0}{\partial z}\right|_{z=z'_-}=-4\pi\ell_B,
\eea
where $\Sigma$ denotes the position of the interface at $z=0$. One finds for $0\leq z\leq z'$
\bea\label{tv0}
\tv_0(z,z',k)=\frac{2\pi\ell_B p_b}{k^2}\left[h_+(z)h_-(z')+\Delta h_-(z)h_-(z')\right],\nonumber\\
\eea
where the delta function is defined as
\be\label{del}
\Delta=\frac{\kappa_b^2\mathrm{csch}^2\left(\kappa_bz_0\right)+(p_b-\eta k)\left[p_b-\kappa_b\coth\left(\kappa_bz_0\right)\right]}
{\kappa_b^2\mathrm{csch}^2\left(\kappa_bz_0\right)+(p_b+\eta k)\left[p_b+\kappa_b\coth\left(\kappa_bz_0\right)\right]},
\ee
with $\eta=\e_m/\e_w$. In the limit $\sigma_s\to 0$ where surface charge induced correlation effects vanish, the potential~(\ref{tv0}) is naturally  reduced to the WC potential of Eq.~(\ref{eq02}). Finally, the electrostatic Green's function in real space follows from Eq.~(\ref{fourker}) as
\be\label{grreal}
v_0(\br,\br')=\int_0^\infty\frac{\mathrm{d}kk}{2\pi}\mathrm{J}_0\left(k|\br_\|-\br'_\||\right)\tv_0(z,z',k),
\ee
and the solution for $z>z'$ is obtained by interchanging in Eq.~(\ref{tv0}) the variables $z$ and $z'$.

The computation of the corrections to the external potential in Eqs.~(\ref{eq342})-(\ref{eq344}) requires the evaluation of the function $\tv_0(z,z',k)$ in the infrared limit, which reads for $0\leq z\leq z'$
\bea\label{eq022}
\tv_0(z,z',k\to 0)&=&\frac{\pi\ell_B}{\kappa_b}e^{-\kappa_bz'}\left\{1+\coth\left[\kappa_b(z'+z_0)\right]\right\}\nonumber\\
&&\times H(z),
\eea
where we introduced the auxiliary function
\bea
H(z)&=&\kappa_bz\hspace{0.5mm}e^{\kappa_bz}-\left(\tilde\Delta+\gamma_c^2(s)\kappa_bz\right)e^{-\kappa_bz}\\
&&+\left\{\left(1-\kappa_bz\right)e^{\kappa_bz}-\left[\tilde\Delta+\gamma_c^2(s)\left(1+\kappa_bz\right)\right]e^{-\kappa_bz}\right\}\nonumber\\
&&\hspace{5mm}\times\coth\left[\kappa_b(z+z_0)\right],
\eea
with
\be
\tilde\Delta=\frac{1}{2}\frac{\gamma_c^2(s)-1}{\gamma_c^2(s)+1}\left[\gamma_c^4(s)-6\gamma_c^2(s)+1\right].
\ee
Interchanging the variables $z$ and $z'$ in Eq.~(\ref{eq022}) yields the function $\tv_0(z,z',k\to 0)$ for $z\geq z'$.

Finally, the self-energy defined in Eq.~(\ref{eq7}) follows from Eq.~(\ref{grreal}) in the form
\bea\label{self}
\delta v_0(z)&=&\ell_B\kappa_b^2\int_0^\infty\frac{\mathrm{d}k}{p_bk}\left\{-\mathrm{csch}^2\left[\kappa_b(z+z_0)\right]\right.\\
&&\left.+\Delta\left(\frac{p_b}{\kappa_b}+\coth\left[\kappa_b(z+z_0)\right]\right)^2e^{-2p_bz}\right\}.\nonumber
\eea
As expected, the potential Eq.~(\ref{self}) tends in the limit $\kappa_b\to0$ to the expression derived in Ref.~\cite{netzcoun} for the asymmetrically partitioned counterion-only system. Furthermore, we note that the potential~(\ref{self}) is similar in form to the one derived in Ref.~\cite{1loop} for a symmetrical salt partition around a charged planar interface. However, due to the asymmetry of the salt distribution in our system as well as the presence of a dielectric discontinuity, the $\Delta$ function in Eq.~(\ref{self}) has a more complicated form and the potential~(\ref{self}) does not posses mirror symmetry with respect to the interface located at $z=0$. Since we will exclusively investigate in this work ion densities in the water medium $z>0$, we do not report here the expression for the ionic self energy in the membrane medium $z<0$.

\subsection{WKB approximation for neutral slit pores}
\label{wkb}
We will explain in this part the solution of the SC equation~(\ref{eq6}) within a WKB approximation in the neutral pore limit $\sigma(z)=0$ where the external potential vanishes, i.e. $\phi(z)=0$. The calculation presented here is an extension of a previous WKB solution of Eq.~(\ref{eq6}) for a single dielectric interface~\cite{japan} to the more complicated case of a slit nanopore. Unlike the approach of Section~\ref{anan} based on a perturbative expansion of the formal solution of SC equations, the WKB approximation has the advantage of being a non-perturbative approach. Consequently, while moving from the bulk towards the pore wall, it will be shown below that the WKB solution can interpolate between the DH regime with $\kappa(z\to\infty)=\kappa_b$ and the dilute electrolyte regime with $\kappa(z\to0)=0$ in a self-consistent way.

First of all, by injecting the Fourier expansion Eq.~(\ref{fourker}) into Eq.~(\ref{eq6}), one gets
\be
\label{eqDHSCfr}
\left[-\partial_z\epsilon(z)\partial_z+\epsilon(z)p^2(z)\right]\tv(z,z',k)=\frac{e^2}{k_BT}\delta(z-z'),
\ee
where we introduced the function $p(z)=\sqrt{k^2+\kappa^2(z)}$ whose $k$-dependence is implicit. Furthermore, we defined a local screening parameter in the form
\be\label{saltSC}
\kappa^2(z)=\kappa_b^2e^{-V_w(z)-\frac{q^2}{2}\delta v(\br,\br)}.
\ee
In the present work, we will exclusively need the Green's function associated with source particles within the slab, i.e. $0\leq z'\leq d$. Without making any approximation yet, the general solution to the second order differential equation~(\ref{eqDHSCfr}) in the slit geometry can be expressed in terms of its two independent homogeneous solutions $h_\pm(z)$ in the form
\begin{widetext}
\be\label{totsol}
\tv(z,z',k)=c_1e^{kz}\theta(-z)+\left[c_2h_+(z)+c_3h_-(z)\right]\theta(z)\theta(z'-z)+\left[c_4h_+(z)+c_5h_-(z)\right]\theta(d-z)\theta(z-z')
+c_6e^{-kz}\theta(z-d),
\ee
where the coefficients $c_i$ are integration constants to be found by imposing the boundary conditions~(\ref{boun1})-(\ref{boun3}) at the interfaces $\Sigma=0$ and $\Sigma=d$. After some long algebra, we get
\be\label{genv}
v(\br,\br')=\ell_B\int_0^\infty\mathrm{d}kk\frac{G(z')}{F(z')}\mathrm{J}_0\left(k|\br_\|-\br'_\||\right)
\left\{\left[\e_w h_-'(0)-\e_mkh_-(0)\right]h_+(z)-\left[\e_w h_+'(0)-\e_mkh_+(0)\right]h_-(z)\right\},
\ee
where
\be
G(z)=\e_w\left[h_+(z)h_-'(d)-h_-(z)h_+'(d)\right]+\e_mk\left[h_+(z)h_-(d)-h_-(z)h_+(d)\right]
\ee
and
\bea
F(z)&=&\left[h_-(z)h_+'(z)-h_+(z)h_-'(z)\right]\left\{\e_w^2\left[h_+'(0)h_-'(d)-h_-'(0)h_+'(d)\right]
+\e_m^2k^2\left[h_-(0)h_+(d)-h_+(0)h_-(d)\right]\right.\\
&&\hspace{4.5cm}\left.+\e_w\e_mk\left[h_+'(0)h_-(d)-h_-'(d)h_+(0)+h_-(0)h_+'(d)-h_+(d)h_-'(0)\right]\right\}.\nonumber
\eea
\end{widetext}
The formal solution~(\ref{genv}) can be actually used in order to solve Eq.~(\ref{eq6}) exactly by evaluating the functions $h_\pm(z)$ from a numerical solution of Eqs.~(\ref{eqDHSCfr}) and~(\ref{saltSC}) in a two dimensional lattice $(k,z)$  by iteration. We will however leave the exploration of this idea for a future work, and evaluate Eq.~(\ref{genv}) within the WKB approximation where the homogeneous solutions of Eq.~(\ref{eqDHSCfr}) read~\cite{japan}
\be\label{homsol}
h_\pm(z)=\left[p(z)\right]^{-1/2}\mathrm{exp}\left[\pm\int_0^z\mathrm{d}z'p(z')\right].
\ee
Furthermore, in order to simplify the numerical task, we will limit ourselves to the case $\e_m<\e_w$ characterized by a continuously varying $\kappa(z)$ on the interval $0\leq z\leq d$ and vanishing ion densities on both sides of the interfaces, i.e.
\be
\lim_{z\to 0^{\pm}}\rho(z)=\lim_{z\to d{\pm}}\rho(z)=0.
\ee
\begin{figure*}
(a)\includegraphics[width=0.46\linewidth]{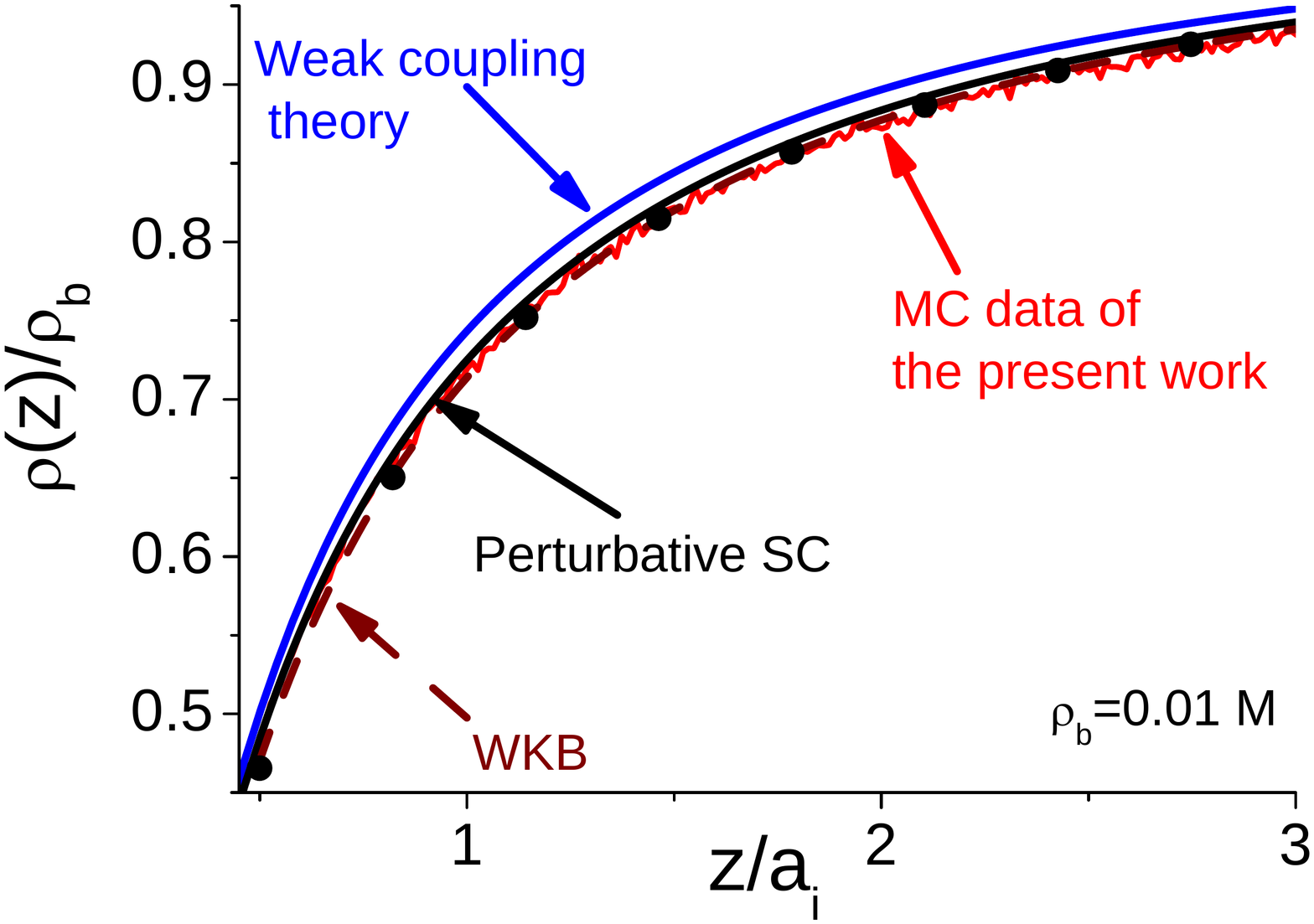}
(b)\includegraphics[width=0.46\linewidth]{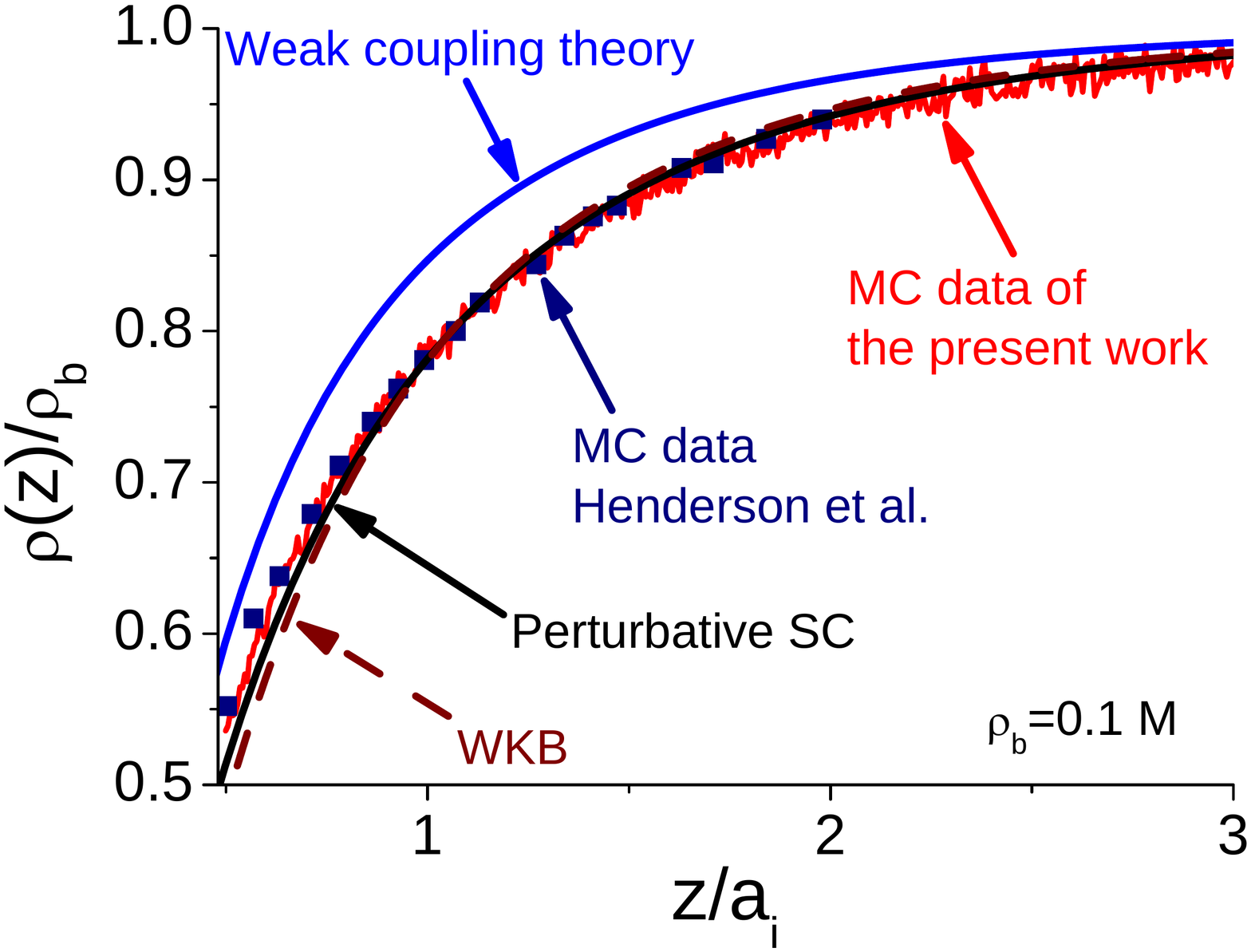}
(c)\includegraphics[width=0.46\linewidth]{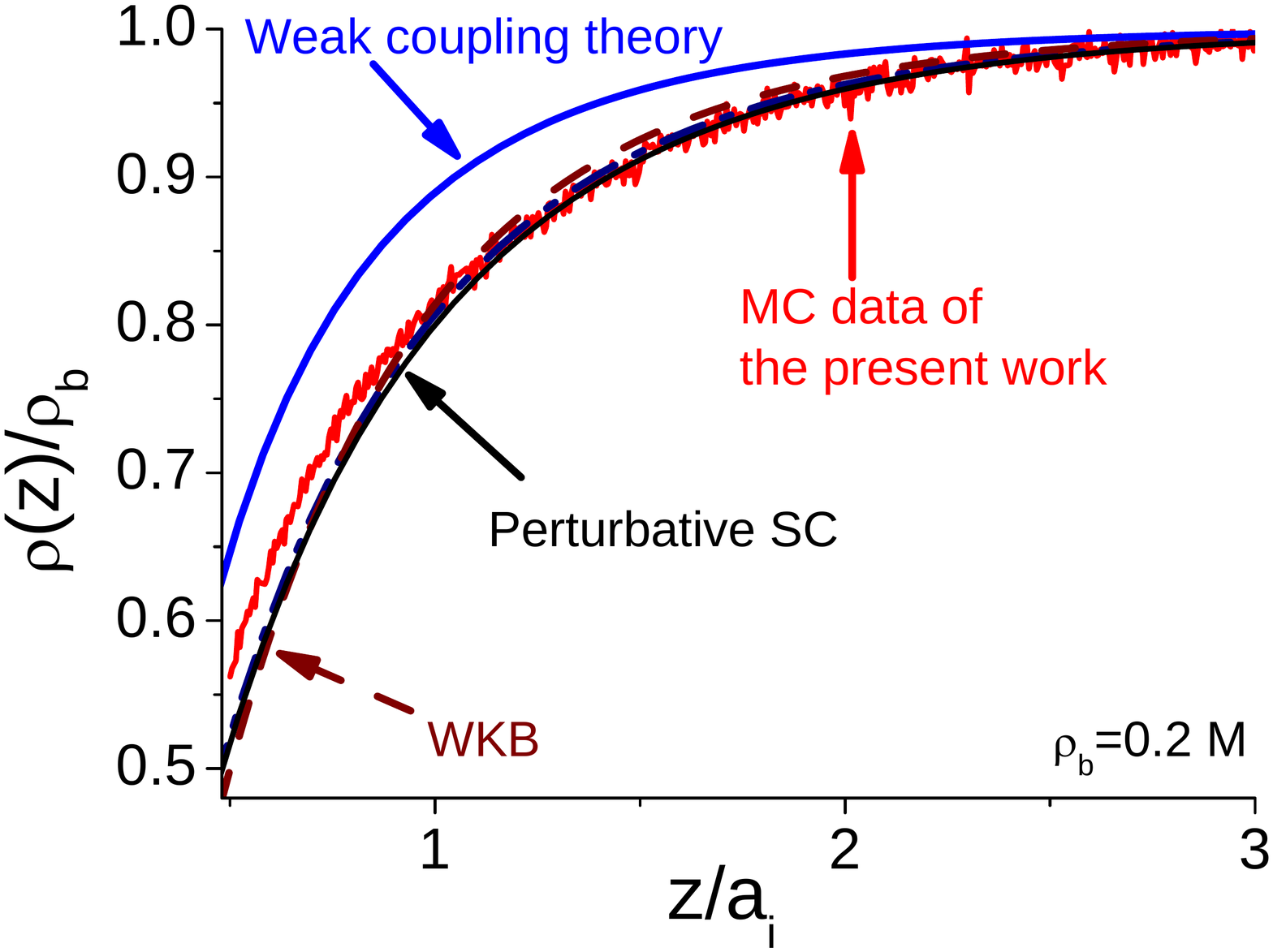}
(d)\includegraphics[width=0.46\linewidth]{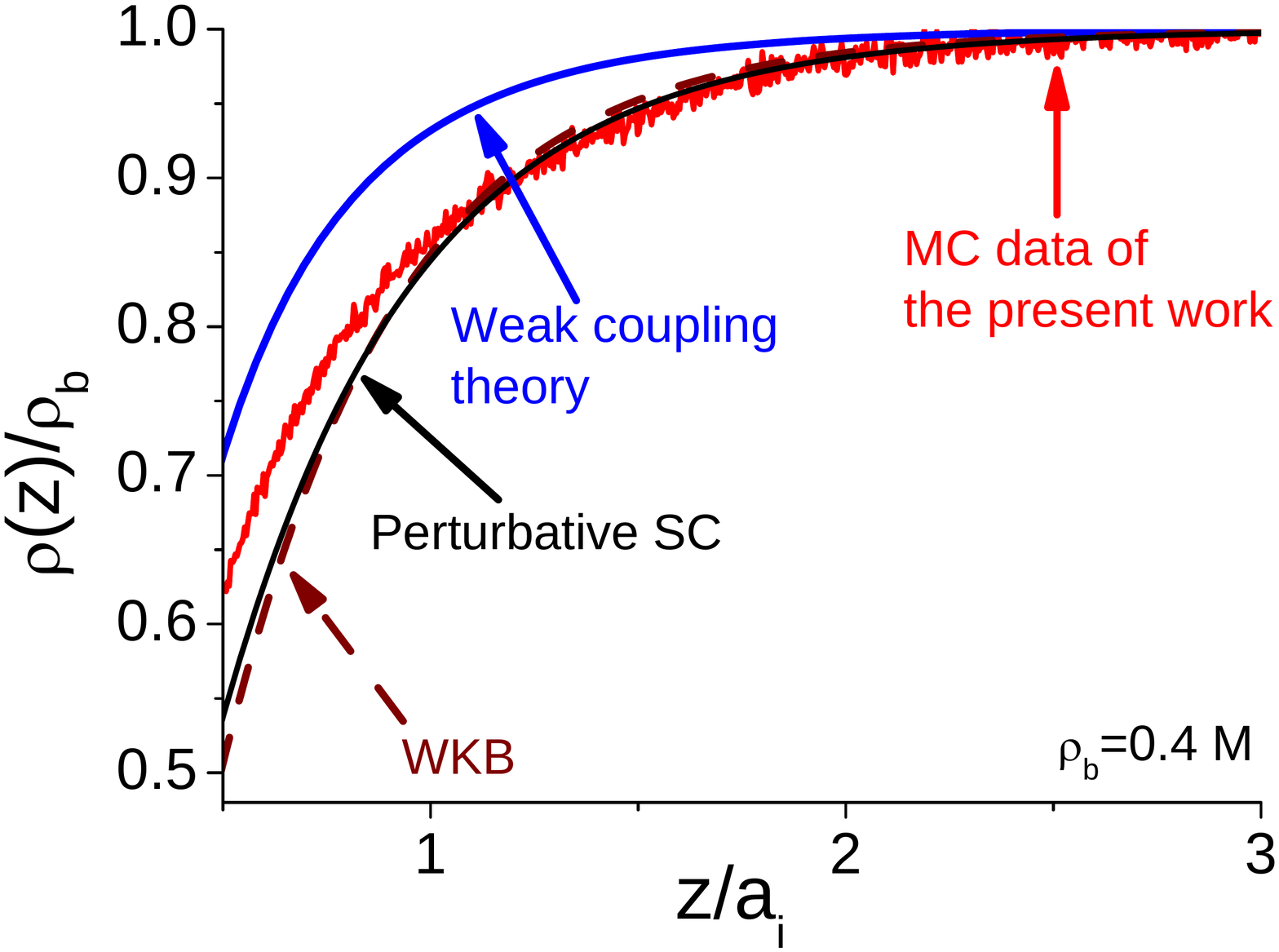}
\caption{(Color online) Ion density profiles at the dielectric interface against the distance from the surface with $\e_m=1$, $\e_w=80$, and ion diameter $a_i=4.25$ {\AA} at the bulk ion concentration (a) $\rho_b=0.01$ M, (b) $\rho_b=0.1$ M, (c) $\rho_b=0.2$ M, and (d) $\rho_b=0.4$ M. The red lines are our MC simulation data, the blue lines the WC theory, and the dashed brown and black lines are respectively the WKB and  the second order perturbative solutions of the SC equation~(\ref{eq6}). The black dots in (a) denote the iterative solution of the closure relations Eqs.~(\ref{saltSC}) and~(\ref{kerself}) explained in Appendix~\ref{persol}, while the blue squares in (b) are the simulation data from Ref.~\cite{hend1}. The dashed blue line in (c) marks the third order perturbative solution of SC equations.}
\label{fig1}
\end{figure*}
After some long but straightforward algebra, one obtains the kernel in the form
\be\label{kertot}
v(\br,\br')=v_b(\br,\br')+v´_d(\br,\br'),
\ee
where the bulk part is given by
\bea\label{kerbulk}
v_b(\br,\br')&=&\ell_B\int_0^\infty\frac{\mathrm{d}kk}{\sqrt{p(z)p(z')}}\mathrm{J}_0\left(k|\br_\|-\br'_\||\right)\nonumber\\
&&\hspace{1.6cm}\times e^{-\left|I(z,z')\right|}
\eea
and the dielectric contribution reads
\bea\label{kerdi}
v_d(\br,\br')&=&\ell_B\Delta_0\int_0^\infty\frac{\mathrm{d}kk}{\sqrt{p(z)p(z')}}\frac{\mathrm{J}_0\left(k|\br_\|-\br'_\||\right)}
{1-\Delta_0^2e^{-2I(0,d)}}\\
&&\hspace{1.6cm}\times\left\{e^{-I(0,z)-I(0,z')}+e^{-I(z,d)-I(z',d)}\right.\nonumber\\
&&\hspace{2.2cm}\left.+2\Delta_0e^{-2I(0,d)}\cosh\left[I(z,z')\right]\right\}.\nonumber
\eea
We defined in Eqs.~(\ref{kerbulk}) and~(\ref{kerdi}) the integral
\be
I(z_1,z_2)=\int_{z_1}^{z_2}\mathrm{d}\tilde{z}p(\tilde{z})
\ee
and introduced the dielectric jump function
\be
\Delta_0=\frac{\e_w-\e_m}{\e_w+\e_m}.
\ee

It is interesting to note that the bulk part of the electrostatic Green's function Eq.~(\ref{kerbulk}) is not of the DH form $v_{DH}(\br,\br')=\ell_B/re^{-\kappa_br}$, with $r=|\br-\br'|$ the interionic distance. The difference between both potentials is clearly due to the breaking of the spherical symmetry by the non-uniform ionic screening, an interfacial effect absent in the DH theory. Moreover, Eq.~(\ref{kerbulk}) shows that by approaching the dielectric surface where the ion density vanishes, the potential~(\ref{kerbulk}) tends to the usual Coulomb law, i.e. $\lim_{z,z'\to0^+}v_b(\br,\br')=\ell_B/r$. In the next part, we will take advantage of the ability of the WKB solution to self-consistently interpolate between the dilute limit and the DH regime in order to evaluate the asymptotic small distance limit of polymer-interface interactions.

For the computation of the local ion densities, we exclusively need the self energy of ions defined in Eq.~(\ref{eq7}), which reads
\bea\label{kerself}
&&\delta v(\br,\br)=\delta v(z)=\ell_B\left[\kappa_b-\kappa(z)\right]\\
&&+\ell_B\Delta_0\int_0^\infty\frac{\mathrm{d}kk}{p(z)}
\frac{e^{-2I(0,z)}+e^{-2I(z,d)}+2\Delta_0e^{-2I(0,d)}}{1-\Delta_0^2e^{-2I(0,d)}}.\nonumber
\eea
We notice that the first term on the rhs of this equation is associated with the local variation of the ionic cloud around the source ion, and since $\kappa(z)<\kappa_b$ for $\e_m<\e_w$, this solvation energy positively adds to the image charge repulsion contribution, i.e. the second term on the rhs of Eq.~(\ref{kerself}).

The relations~(\ref{saltSC}) and~(\ref{kerself}) form a set of closure equations that has to be solved by iteration on a bidimensional lattice $(k,z)$. The iterative approach consists in injecting into Eq.~(\ref{saltSC}) the ionic self energy obtained from the WC potential Eq~(\ref{eq01}), which yields an updated screening function $\kappa_1(z)$. One then evaluates the new self energy profile from Eq.~(\ref{kerself}) with $\kappa_1(z)$, and the iterative cycle continues until self-consistency is achieved. We note that this iteration method was used in Ref.~\cite{japan} to compute an analytical expression for the ionic self energy at single dielectric interfaces with $\e_m=0$. An extension of this calculation to finite values of $\e_m$ is presented in Appendix~\ref{persol}. Before concluding, we also note that for simple interfaces, the ionic self energy~(\ref{kerself}) reduces to the expression derived in Ref.~\cite{japan},
\be\label{eq10}
\delta v(z)=\ell_B\left[\kappa_b-\kappa(z)\right]+\ell_B\Delta_0\int_0^\infty\frac{\mathrm{d}kk}{p_c(z)}e^{-2I(0,z)}.
\ee
\begin{figure}
\includegraphics[width=1.15\linewidth]{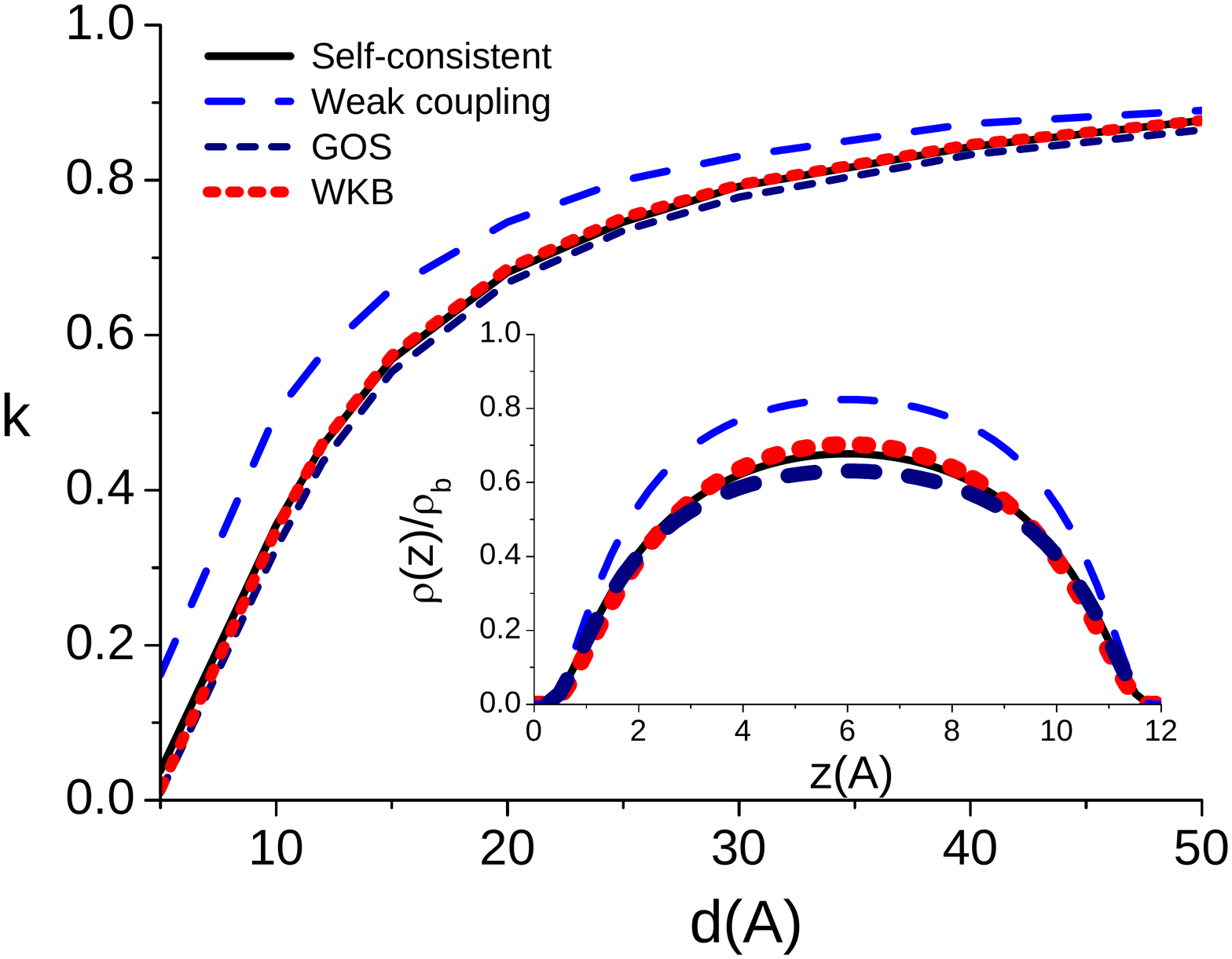}
\caption{(Color online) Pore averaged ion densities in slit pores against the pore size for $\rho_b=0.1$ M, $\e_m=1$, and $\e_w=80$. The light blue line is the WC theory, the black line is the SC theory, the red line is the WKB solution, and the dark blue line denotes the restricted variational scheme of Ref.~\cite{PRE}. The inset displays local ion densities in the pore for the same model parameters.}
\label{fig2}
\end{figure}

\section{Results}
\label{res}
\subsection{Neutral interfaces}

We will first determine in this part the salt concentration range where the SC equation~(\ref{eq6}) remains quantitatively accurate for electrolyte systems in contact with neutral dielectric interfaces separating the solvent part with ions from a membrane region free of ions. This geometry is relevant to the water-air interface as well as to membrane nanopores characterized by strong ionic confinement effects. The computation schemes developed in the previous parts will be then applied within this range to slit nanopores and polymer-interface systems in order to test the validity of the DH theory and a restricted variational approach~\cite{PRE} relevant to experimental nanofiltration studies~\cite{Szymczyk}.

\subsubsection{Ion densities}

We establish in this part the validity domain of the SC equation~(\ref{eq6}) for neutral single interfaces and slit nanopores by comparing the predictions of the theoretical schemes developed in the previous parts with MC simulation data for ion densities. The details of our numerical simulations can be found in Appendix~\ref{mcdetails}. We note that in order to be able to compare our simulation results with previous MC simulation data from Ref.~\cite{hend1}, we chose the diameter of ions  as $a_i=4.25$ {\AA}, and the dielectric permittivity of the membrane and the water respectively as $\e_m=1$ and $\e_w=80$. The comparison of our MC data in Fig.~\ref{fig1}(b) with the numerical results of Ref.~\cite{hend1} shows the good agreement between the two simulation results.

The  comparison of the WC theory in Fig.~\ref{fig1}(a) with MC data shows that the concentration $\rho_b=0.01$ M marks the boundary of the ion density range where the WC theory starts to deviate from the simulation result, although even at this concentration, it exhibits a reasonably good quantitative agreement with the MC data and the SC theory. We note that this density corresponds to an electrostatic coupling parameter $\Gamma=\kappa_b\ell_B\simeq0.2$. For larger bulk concentrations in Figs.~\ref{fig1}.b-d where this deviation becomes more pronounced, one sees that the WC theory systematically overestimates the ion density.
Within the restricted variational theory of Refs.~\cite{hatlo,hatlorev}, this overestimation was shown to originate from the unability of the WC image potential of Eq.~(\ref{eq02}) to account for the reduction of the  ionic screening at the interface. The transparency of the present method allows to confirm this conclusion at an analytical level. Indeed, the first order correction to the WC Green's function in Eq.~(\ref{eq339}) shows with Eq.~(\ref{nper}) that due to the interfacial ion deficiency $\delta n_0(z)<0$, the positive correction to the image potential $\delta v_{10}(z)>0$ increases the amplitude of the WC potential, leading to a stronger dielectric exclusion at the interface.

An inspection of Figs.~\ref{fig1}.b-d shows that the predictions of the WKB and the perturbative solutions of the SC equation~(\ref{eq6}) both exhibit a good agreement with the simulation data up to $\rho_b=0.2$ M, thus improving the quantitative accuracy of the WC theory approximately by one order of magnitude in ionic strength. Hence, the deviation of the SC theory from the simulation data taking place at $\rho_b=0.2$ M (or $\Gamma\simeq1.0$) establishes this value as the characteristic density where the quantitative accuracy of the SC theory breaks down. We note however that it is unclear whether the failure results from electrostatic correlation effects, or excluded volume effects not included in the SC theory that start to set on. This point can be enlightened in future by solving the extended SC equations of Ref.~\cite{jstat} that can account for ionic excluded volume effect.

Before concluding the discussion of the ionic partition at single dielectric interfaces, we would like to note two points. First of all, we show in Fig.~\ref{fig1} that the approximative solution of the closure relations~(\ref{saltSC}) and~(\ref{kerself}) by iteration explained in Appendix~\ref{persol} fits very well their numerical solution in the dilute regime $\rho_b=0.01$ M. This observation is useful since this approximative solution will be used in the next section for an analytical evaluation of the interaction of a rigid polymer with the dielectric interface. Then, we emphasize that the perturbative SC solutions reported in Fig.~\ref{fig1} are obtained at the second order perturbative level introduced in Sec.~\ref{anan}. To ascertain the convergence of our perturbative scheme, we also reported in Fig.~\ref{fig1}(c)  the third order calculation explained in Appendix~\ref{3th} (see the dashed dark blue curve). A careful inspection shows that the third order result is hardly distinguishable from the second order result, which confirms that for neutral single dielectric interfaces, the second order perturbative solution is sufficient within the validity domain of the SC equation~(\ref{eq6}).

We now illustrate in Fig.~\ref{fig2} the ion densities and partition functions $k=\lan\rho(z)/\rho_b\ran_p$ in slit pores with $\e_m=1$ and $\e_w=80$, in contact with a bulk reservoir of ionic concentration $\rho_b=0.1$ M. The plots compare the results obtained from the WC theory, the perturbative and WKB solutions of the SC approach, and the generalized Onsager-Samaras (GOS) approach introduced in Ref.~\cite{PRE}. We note that the model parameters in this figure were chosen in such a way that the results obtained from the second and the third order perturbative solutions of the SC equation~(\ref{eq6}) are practically superimposed, thus guaranteeing the convergence of the perturbative solution. An inspection of the main plot and the inset of Fig.~\ref{fig2} shows that as in the single interface case, the WC theory underestimates the dielectric exclusion, while the WKB and the perturbative solutions of the SC theory are again very close to each other. Moreover, one notices in the inset of the same figure that the GOS formalism is able to accurately reproduce the ion density close to the interface, but slightly overestimates the dielectric exclusion in the mid-pore area. However, we see in the main plot that despite this weak discrepancy, the GOS approach remains very accurate in estimating the partition coefficients over a large interval of pore thickness. This observation might indeed explain the success of GOS-like solutions of the SC equation~(\ref{eq6}) frequently used in artificial nanofiltration studies in order to estimate exprimental salt rejection rates~\cite{yarosch,YarII,Szymczyk}.

\subsubsection{Polymer-surface interactions}

In this section, we will evaluate within the WKB approach introduced in Section~\ref{wkb}  the energetic cost to drive a rigid polymer from the bulk reservoir to the proximity of the interface. If the surface charge density of the polymer $\tau$ is weak enough so that it does not significantly affect the interfacial ion densities, the electrostatic energy of the polymer located at $(y=0,z)$ can be obtained from the relation~\cite{netzdh}
\be\label{frtot}
F(z)=\int\frac{\mathrm{d}\br_1\mathrm{d}\br_2}{2}\sigma(\br_1)v(\br_1,\br_2)\sigma(\br_2),
\ee
where the electrostatic Green's function is given by Eq.~(\ref{kertot}), and the linear charge density of the polymer is
\be\label{chdis}
\sigma(\br')=\tau\delta(y')\delta(z'-z).
\ee
The net energetic cost for bringing the polymer from the bulk to the dielectric surface located at $z=0$ is given by
\be
\Delta f(z)=\frac{1}{L}\left[F(z)-F_b\right],
\ee
where $L$ is the polymer length, and the total electrostatic energy of the polymer in the bulk electrolyte reads $F_b=\lim_{z\to\infty}\lim_{d\to\infty}F(z)$. We note that within the DH theory, the same energy density was derived in Ref.~\cite{netzdh} in the form
\be\label{frdh}
\Delta f_{DH}(z)=\frac{\ell_B\tau^2}{2}\int_{-\infty}^{\infty}\frac{\mathrm{d}k_y}{p_b}\Delta_b e^{-2p_bz},
\ee
where $p_b$ and $\Delta_b$ are respectively given by Eqs.~(\ref{pb}) and~(\ref{deltab}). In the asymptotic limit of small polymer surface separations, Eq.~(\ref{frdh}) was also shown to reduce to a simple logarithmic law,
\be
\Delta f_{DH}(z)\simeq\Delta_0\ell_B\tau^2\ln\left(1/\kappa_bz\right).
\ee
\begin{figure}
\includegraphics[width=1.05\linewidth]{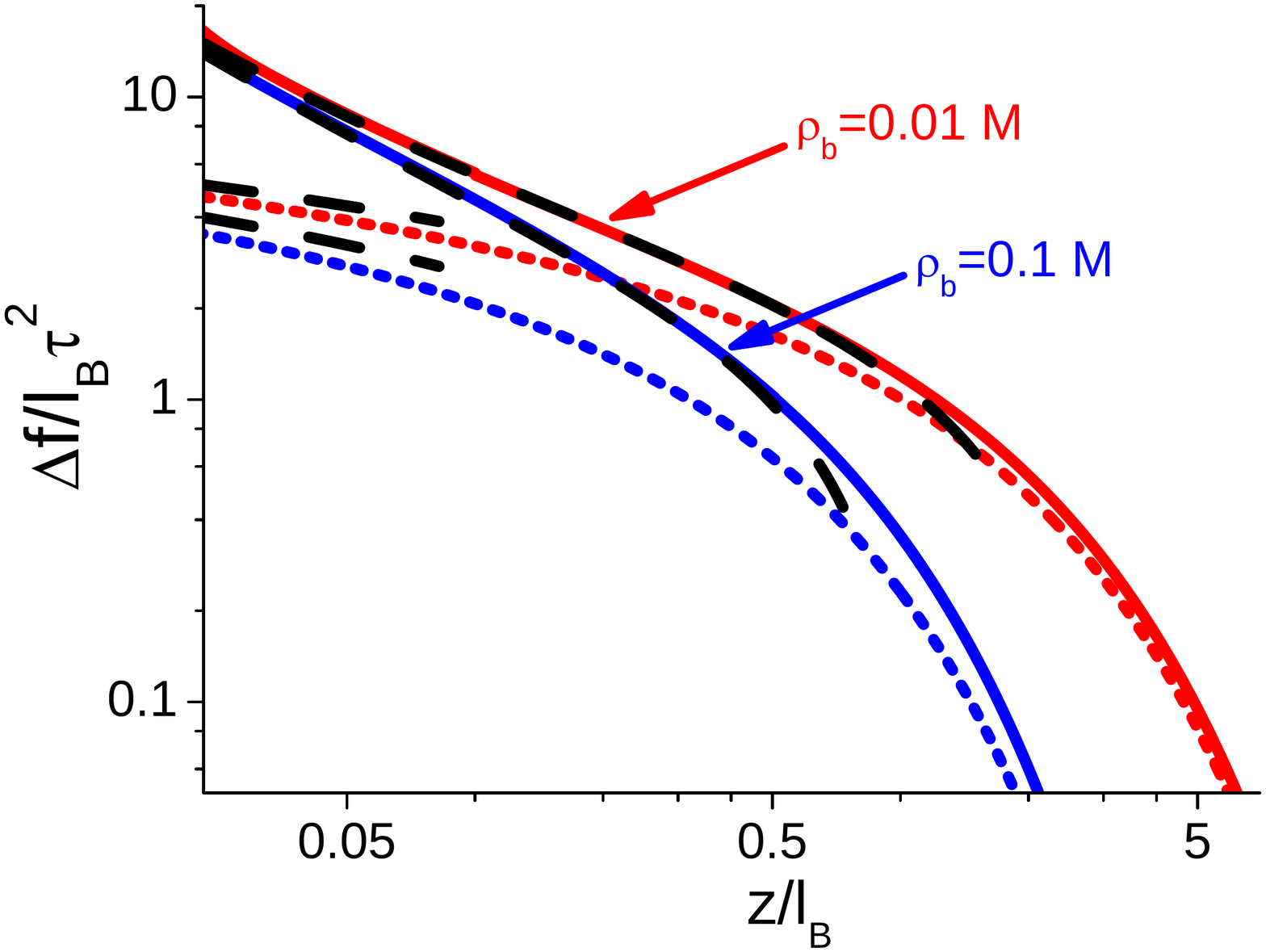}
\caption{(Color online) Reduced polymer-interface free energy against the distance from the dielectric surface for the membrane permittivity $\e_m=1$, and salt concentrations $\rho_b=0.01$ (red curves) and 0.1 M (blue curves). The solid lines are from the WKB theory and the dashed lines denote the prediction of the DH theory. The black dashed lines mark for each curve the asymptotic small $z$ behavior.}
\label{fig3}
\end{figure}

Evaluating the integrals in Eq.~(\ref{frtot}) with the Green's function Eq.~(\ref{kertot}) in the limit $d\to\infty$ and the polymer charge distribution of Eq.~(\ref{chdis}), we obtain the free energy profile as
\be\label{fr1int}
\Delta f(z)=\ell_B\tau^2\ln\left[\kappa_b/\kappa(z)\right]+\ell_B\tau^2\frac{\Delta_0}{2}\int_{-\infty}^{\infty}\frac{\mathrm{d}k_y}{p(z)}e^{-2I(0,z)}.
\ee
We emphasize that since the WKB solution was derived in Sec.~\ref{wkb} for the case of a
dielectric discontinuity where the ion density vanishes at the interface, the relation~(\ref{fr1int}) is valid in the permittivity range $\e_m<\e_w$.

It is seen that Eq.~(\ref{fr1int}) is composed of a local ionic solvation part absent at the DH level, and a second term associated with the dielectric jump at the interface. The solvation term that depends logarithmically on the ionic density accounts for the spatial variations of the ionic cloud density around the charged polymer strand. Since the screening of the polymer charge lowers its free energy, this contribution acts as an additional force pushing the polymer towards the bulk area where the ionic density is maximum.

DH and SC free energy profiles of Eq.~(\ref{frdh}) and~(\ref{fr1int}) are illustrated in Fig.~\ref{fig3} for the membrane permittivity $\e_m=1$, and salt concentrations $\rho_b=0.01$ and $0.1$ M. We first notice that the free energy barrier evaluated within the SC theory is significantly larger than the prediction of the WC theory. Then, one notices that in the close neighborhood of the interface, the same barrier is characterized by a regime that is clearly independent of the bulk salt concentration.

In order to elucidate these two points, we will evaluate the asymptotic small $z$ limit of  Eq.~(\ref{fr1int}) within the perturbative approach explained in Appendix~\ref{persol}. To this end, we inject into Eq.~(\ref{fr1int}) the screening function Eq.~(\ref{saltSC}) evaluated at the first order iterative level with the density profile Eq.~(\ref{den1}),
\be
\kappa_1^2(z)=\kappa_0^2(z)\left[1-\frac{q^2}{2}\delta v_1(z)\right],
\ee
where we defined
\bea
\kappa_0^2(z)=\kappa_b^2\mathrm{exp}\left(-\frac{q^2\ell_B\Delta_0}{4z}e^{-2\kappa_bz}\right),
\eea
and expand the result in powers of $\delta v_1(z)$.  At the leading order $O\left(\delta v_1(z)\right)$, we get
\bea\label{fr1int01}
\Delta f(z)&\simeq&\ell_B\tau^2\ln\left[\kappa_b/\kappa_0(z)\right]+\ell_B\tau^2\frac{\Delta_0}{2}\int_{-\infty}^{\infty}\frac{\mathrm{d}k_y}{p_0(z)}e^{-2I_0(0,z)}\nonumber\\
&&+\frac{\ell_Bq^2\tau^2}{4}\delta v_1(z)\\
&&+\frac{\Delta_0\ell_Bq^2\tau^2}{2}\int_0^\infty\frac{\mathrm{d}k_y}{p_0(z)}\left\{\frac{\kappa_0^2(z)}{2p_0^2(z)}\delta v_1(z)\right.\nonumber\\
&&\hspace{2cm}\left.+\int_0^z\mathrm{d}z'\frac{\kappa_0^2(z')}{p_0(z')}\delta v_1(z')\right\}e^{-2I_0(0,z)},\nonumber
\eea
where the subscript $0$ means that the screening function $\kappa(z)$ in the functions $p(z)$ and $I(0,z)$ should be replaced with $\kappa_0(z)$. We will now derive from Eq.~(\ref{fr1int01}) the asymptotic limit of the free energy close to the interface. If we note that the screening function $\kappa(z)$ is very small with respect to $\kappa_b$ in the neighborhood of the surface, we can assume that $p_0(z)=\sqrt{k^2+\kappa^2(z)}$  varies slowly close to the dielectric interface. One can thus take the function $p_0(z)$ out of the integral in $I_0(0,z)$, and carry out the integration over $k_y$. Expanding the result up to the linear order in $z$, one obtains
\bea\label{asym1}
\Delta f(z)&\simeq&\frac{\Delta_0(1+\Delta_0)\ell_B^2\tau^2q^2}{8z}+\Delta_0\ell_B\tau^2\ln\left(1/\kappa_bz\right)\nonumber\\
&&-\gamma\Delta_0\ell_B\tau^2+\frac{q^2}{4}\ell_B^2\tau^2\kappa_b\left(1-\frac{3}{4}\Delta_0^2\right),
\eea
where $\gamma\simeq0.5772$ stands for the Euler's constant. The asymptotic law~(\ref{asym1}) reported in Fig.~\ref{fig3} by dashed black lines is shown to accurately reproduce the behavior of the numerically computed free energy barriers close to the surface. We see in Eq.~(\ref{asym1}) as well as in Fig.~\ref{fig3} that the logarithmic dependence of the energy barrier on $z$ predicted by the WC theory is actually dominated by an unscreened algebraic decay. This algebraic regime associated with the interfacial dielectric exclusion of ions explains the salt free portion of the free energy barrier as well as its strong amplitude observed in Fig.~\ref{fig3}.
\begin{figure}
(a)\includegraphics[width=0.9\linewidth]{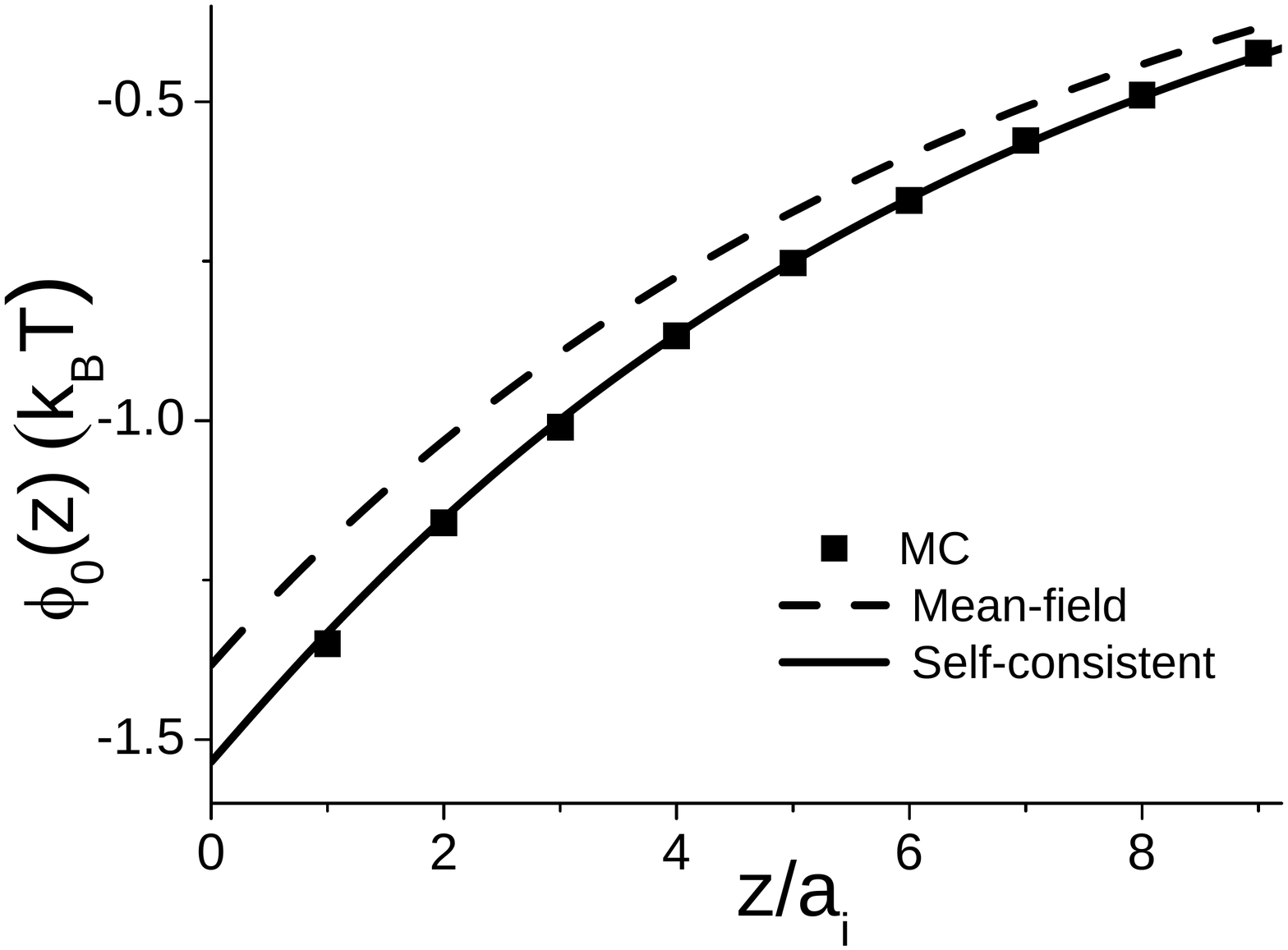}
(b)\includegraphics[width=0.9\linewidth]{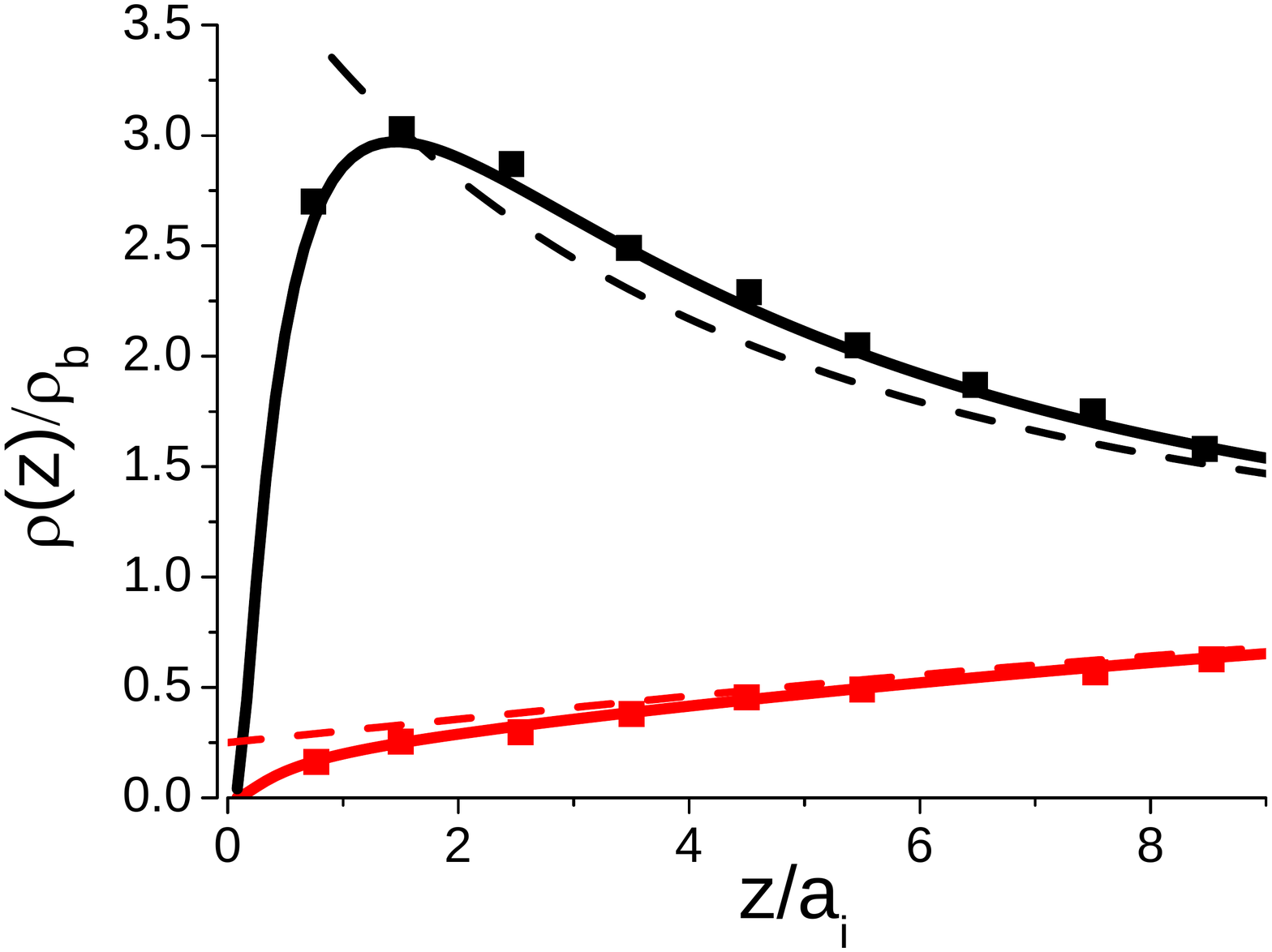}
(c)\includegraphics[width=0.9\linewidth]{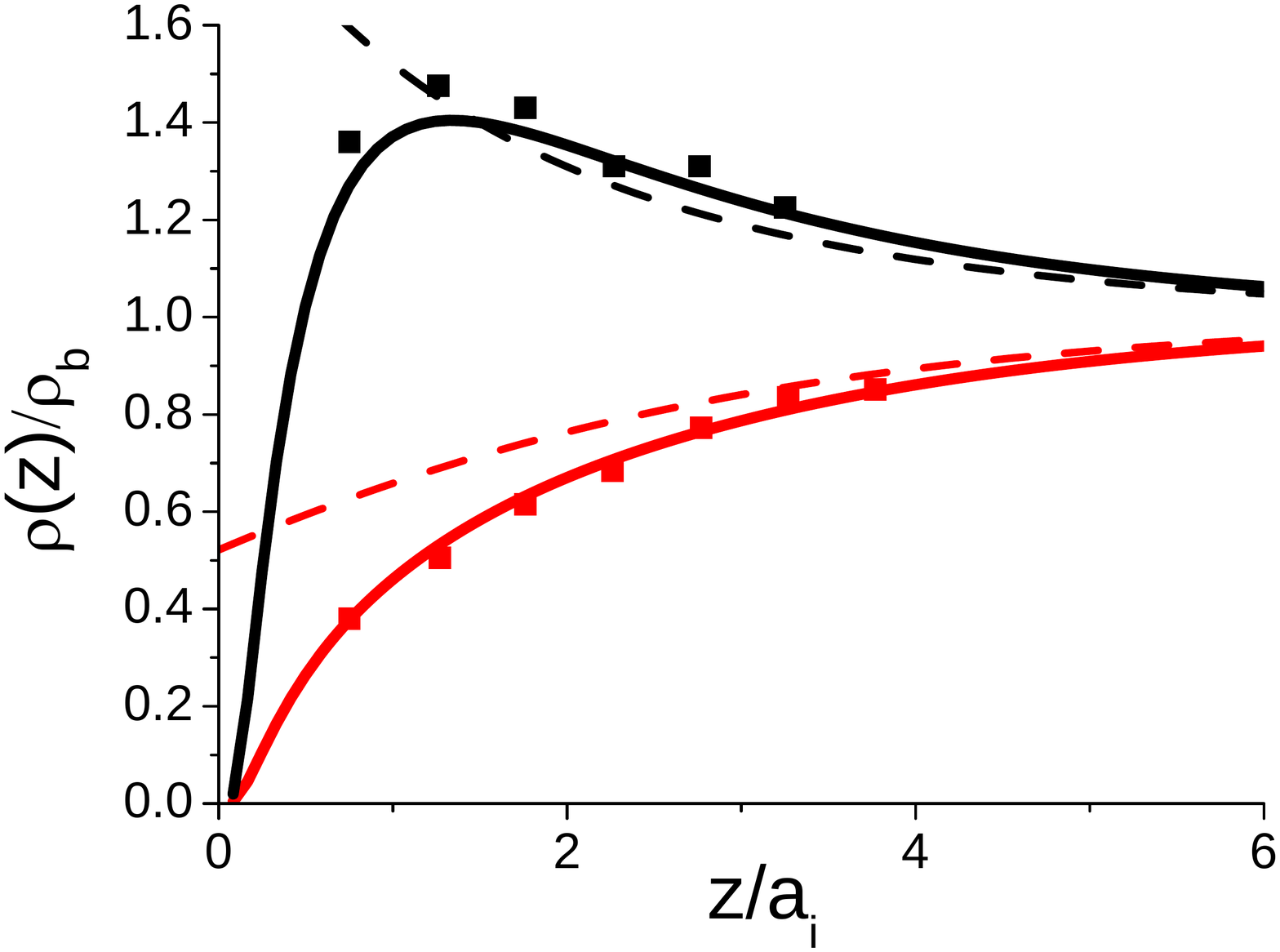}
\caption{(Color online) (a) Electrostatic potential profile for $\rho_b=0.01$ M, and counterion (upper black curves) and coion (lower red curves) densities for $\rho_b=0.01$ M (b) and 0.1 M (c). The solid curves are the SC theory, the dashed lines denote the MF result, and the squares are the MC simulation data from Ref.~\cite{torrie} with ion diameter $a_i=4.25$ {\AA}, $\e_m=1$, and $\e_w=78.5$.}
\label{fig4}
\end{figure}

\subsection{Charged interfaces}

We apply in this part the perturbative SC scheme introduced in section~\ref{anan} to charged single interface systems in contact with a symmetric electrolyte composed of monovalent ions $q=1$. By comparing the theoretical predictions for ion densities and the interfacial electrostatic potential profile with MC simulation data, we will first investigate electrostatic correlation effects at charged dielectric interfaces. The perturbative SC approach presents itself as a useful computational tool particularly in this case of a dielectrically discontinuous system where the one-loop expansion is known to fail~\cite{David}. We will then expand in the limit $\e_m=\e_w$ the perturbative solutions of the SC equations~(\ref{eq5}) and~(\ref{eq6}) introduced in section~\ref{anan} at one-loop order, which will provide us with an analytical one-loop theory of asymmetrically partitioned symmetric electrolyte solutions around charged planar interfaces.

\subsubsection{Correlation effects at charged dielectric interfaces}

We illustrate in Fig.~\ref{fig4}(a) and (b) the prediction of the MF theory and the second order perturbative solution of the SC theory for the electrostatic potential and the ion densities close to a charged interface with the surface charge $\sigma_s=5.5\times10^{-2}$ e $\mathrm{nm}^{-2}$, the bulk ion density $\rho_b=0.01$ M, and the dielectric permittivity values $\e_m=1$ and $\e_w=78.5$. Furthermore, Fig.~\ref{fig4}(c) displays the local ion densities for the same dielectric constants, but with a slightly stronger surface charge $\sigma_s=7.75\times10^{-2}$ e $\mathrm{nm}^{-2}$ and a significantly higher salt concentration $\rho_b=0.1$ M. We also reported for the three plots the MC simulation data from Ref.~\cite{torrie} for ions of diameter $a_i=4.25$ {\AA}.

The first point to be noted in these plots is the good quantitative accuracy of the SC theory in predicting the electrostatic potential profile and the ion densities obtained from MC simulations in the neighborhood of the charged dielectric interface. This is particularly noticeable for the case $\rho_b=0.1$ M where the MF result exhibits a clear disagreement with the simulation data.  The deviation of the MF curves from the MC data can be explained as follows. First of all, the equations~(\ref{eq342}) and~(\ref{sigper}) show that the ionic screening deficiency induced by the dielectric exclusion at the interface gives rise to a positive charge density excess, i.e. $\delta\sigma_s(z)<0$, which in turn results in a negative correction $\psi_{01}(z)<0$ to the MF potential, as can be seen in Fig.~\ref{fig4}(a). Figs.~\ref{fig4}(b) and~\ref{fig4}(c) show that the underestimation of the strength of the electrostatic potential by the MF theory is responsible for an underestimation of the counterion attraction and the coion repulsion by the surface charge for $z>2$ {\AA}. This means that in the presence of a strong dielectric discontinuity between the solvent and the weakly charged substrate, correlation effects reduce the amplitude of the MF level charge separation. Moreover, because the MF theory is unable to account for the charge-image repulsion that dominates the ion-surface charge interaction for $z<2$ {\AA}, both coion and counterion densities are overestimated by the MF theory in the close neighborhood of the interface. However, it will be shown in the next part that in the case $\e_m=\e_w$, this picture is gradually reversed with increasing surface charge.

\subsubsection{One-loop theory of asymmetrically partitioned electrolytes}

We present in this part the one-loop theory of asymmetrically distributed electrolytes around a charged surface that will be shown to directly follow from the perturbative SC scheme developed in section~\ref{anan}. This one-loop calculation that will allow an analytical investigation of electrostatic correlation effects bridges a gap between the DH theory of ions at neutral dielectric interfaces~\cite{netzdh} and the one-loop theory of counterion liquids at charged interfaces~\cite{netzcoun}.

By rescaling first all lengths with the screening parameter according to $\bz=\kappa_b z$ in Eqs.~(\ref{eq339})-(\ref{eq344}) and in Eq.~(\ref{self}), and expanding in the limit $\e_m=\e_w$ the same equations in powers of the electrostatic coupling parameter $\Gamma=\ell_B\kappa_b$, one obtains the following series
\bea\label{1lexp1}
\delta v_0(\bz)&=&\Gamma\bar{\delta v}_0(\bz;s)\\
\label{1lexp2}
\delta v_{nm}(\bz)&=&\sum_{i\geq n+m+1}\Gamma^i\bar{\delta v}_{nm}^{(i)}(\bz;s)\\
\label{1lexp3}
\psi_{nm}(\bz)&=&\sum_{i\geq n+m}\Gamma^i\bar{\psi}_{nm}^{(i)}(\bz;s),
\eea
where the functions under the bar sign depend exclusively on the dimensionless distance $\bar z$ and the parameter $s=\kappa_b\mu$. We note that for $\e_m\neq\e_w$, the self-energy $\delta v_0(z)$ defined in Eq.~(\ref{self}) and appearing in the argument of the exponentials in Eqs.~(\ref{sigper})-(\ref{nper}) becomes singular at $z=0$. Thus, such a loop expansion is valid exclusively in the absence of a dielectric discontinuity~\cite{David}.

The equations~(\ref{1lexp1})-(\ref{1lexp3}) show that the only potentials surviving at the one-loop level are the ion self-energy $\delta v_0(z)$, which is purely on the order $\Gamma$, and the one-loop part of the correction $\psi_{01}(z)$ to the MF external potential $\psi_{1l}(\bz)\equiv\Gamma\bar\psi^{(1)}_{01}(\bz;s)$, which in turn reads
\be
\label{pot1l0}
\psi_{1l}(z)=\rho_bq^4\int_0^\infty\mathrm{d}z_1\tv_0(z,z_1,0)\delta v_0(z_1)\sinh\left[\varphi(z_1)\right].
\ee
A simpler expression for this one-loop potential correction will be given below. Furthermore, by expanding Eq.~(\ref{denper}) up to the order $O(\Gamma)$, the one-loop density follows in the form
\be
\rho^{1l}_{\pm}(\bz)=\rho_{bi}e^{-V_w(\bz)\mp\varphi(\bz)}\left[1-\frac{q^2}{2}\delta v_0(\bz)\mp\psi_{1l}(\bz)\right].
\ee
Expanding now the differential equation~(\ref{eq111}) satisfied by the fluctuating part of the external potential $\psi(\bz)$ at the one-loop order, one obtains
\bea\label{eq1loop}
&&\frac{\partial^2\psi_{1l}(\bz)}{\partial \bz^2}-e^{-V_w(\bz)}\cosh\left[\varphi(\bz)\right]\psi_{1l}(\bz)\nonumber\\
&=&-\frac{q^2}{2}e^{-V_w(\bz)}\delta v_0(\bz)\sinh\left[\varphi(\bz)\right].
\eea
The one-loop correction to the external potential was computed in Ref.~\cite{1loop} by solving the equivalent of the equation~(\ref{eq1loop}) for the charged interface system of a symmetric salt distribution, which is formally equivalent to the calculation above since we note that the inversion of Eq.~(\ref{eq1loop}) directly yields the expression~(\ref{pot1l0}). Furthermore, by integrating Eq.~(\ref{eq1loop}) from $z=-\infty$ to $z=+\infty$ and using the expression~(\ref{pot1l0}), one can show as in Ref.~\cite{1loop}  that the one-loop solution~(\ref{pot1l0}) automatically satisfies the global electroneutrality condition, that is $q\int_0^\infty\mathrm{d}z\left[\rho^{1l}_+(z)-\rho^{1l}_-(z)\right]=\sigma_s$. In other words, as stressed in Ref.~\cite{netzcoun} for the case of the inhomogeneous counterion liquid in contact with a charged interface, the total charge density gets a vanishing contribution from the one-loop theory. Moreover, we note that the one-loop potential $\varphi(z)+\psi_{1l}(z)$ satisfies the Gauss law since one finds from Eq.~(\ref{pot1l0}) that the one-loop correction has a vanishing derivative on the surface, i.e. $\psi'_{1l}(0^+)=0$.

\begin{figure}
(a)\includegraphics[width=1.1\linewidth]{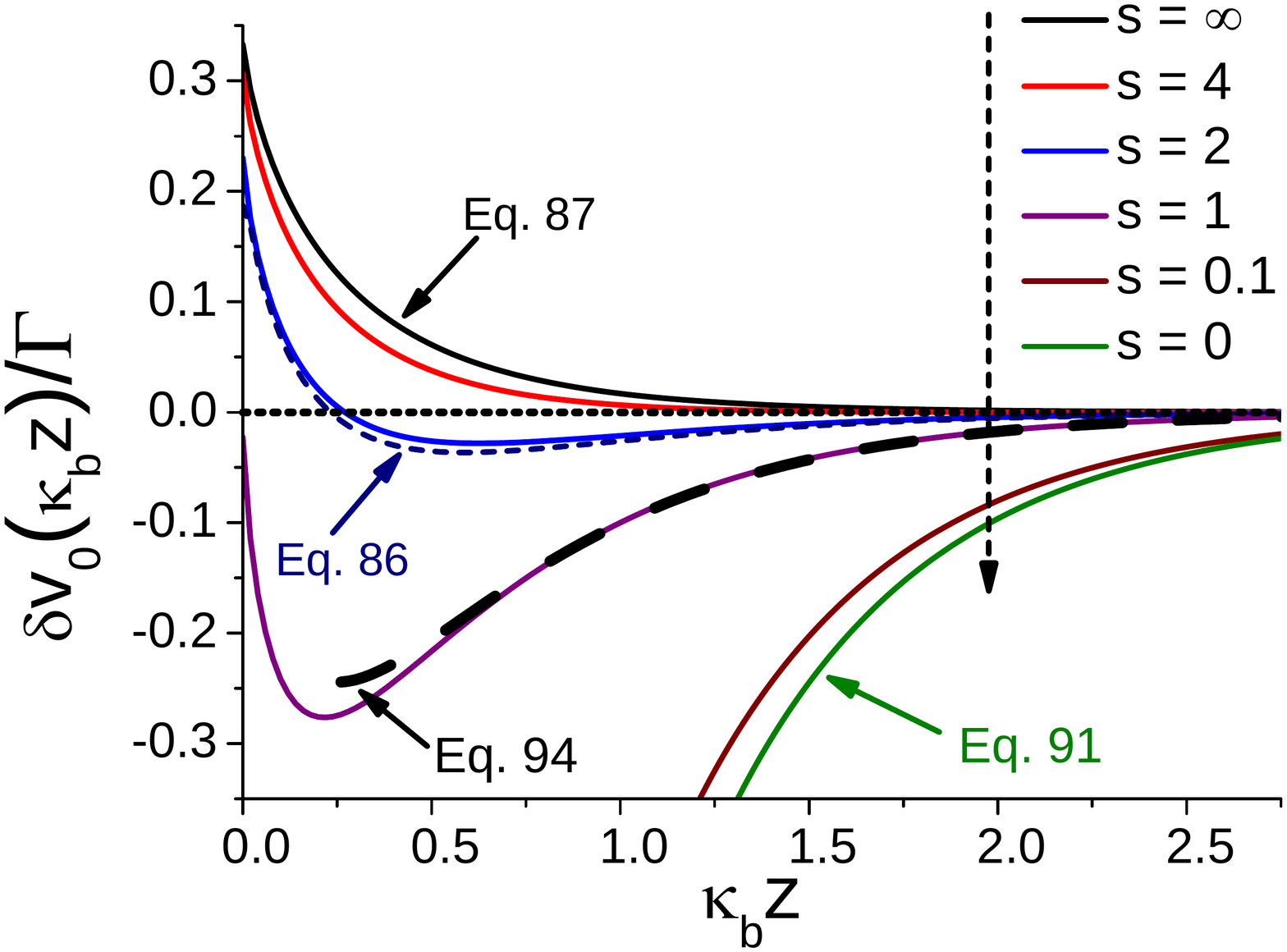}
\caption{(Color online) (a) Ionic self energy renormalized by the coupling parameter $\Gamma$ against the reduced distance from the interface $\kappa_bz$ for different values of the parameter $s$. See text for details.}
\label{fig5}
\end{figure}

We now note that performing the variable transformation $k\to u= p_b/\kappa_b$ in the integral of Eq.~(\ref{self}), the expression for the ionic self-energy can be recast in a more manageable form
\bea\label{self2}
\delta v_0(\bz)&=&\Gamma\int_1^\infty\frac{\mathrm{d}u}{u^2-1}\left\{-\mathrm{csch}^2\left[\bz-\ln\gamma_c(s)\right]\right.\\
&&\left.+\bd\left(u+\coth\left[\bz-\ln\gamma_c(s)\right]\right)^2e^{-2\bz u}\right\},\nonumber
\eea
where we introduced the function
\be
\bd=\frac{1+\left(u-\sqrt{u^2-1}\right)s\left(su-\sqrt{s^2+1}\right)}{1+\left(u+\sqrt{u^2-1}\right)s\left(su+\sqrt{s^2+1}\right)}.
\ee
In the regime of weak surface charges or high salt concentrations corresponding to a large value of $s$, the integral in Eq.~(\ref{self2}) can be analytically evaluated by expanding the integrand in powers of $1/s$. At the leading order, one finds
\be\label{dvLs}
\delta v_0(\bz)=\delta v_{0n}(\bz)+s^{-2}\delta v_{0c}(\bz)+O\left(s^{-4}\right),
\ee
where the vanishing surface charge part was previously derived in Ref.~\cite{PRE} in the form
\be\label{dvLsN}
\frac{\delta v_{0n}(\bz)}{\Gamma}=\frac{(1+\bz)^2}{2\bz^3}e^{-2\bz}-\frac{1}{\bz}\mathrm{K}_2(2\bz),
\ee
and the surface charge contribution is now given by
\bea\label{dvLsC}
\frac{\delta v_{0c}(\bz)}{\Gamma}&=&2\left(\frac{1}{\bz}-1-e^{-2\bz}\right)\mathrm{K}_0(2\bz)\\
&&+\frac{2}{\bz^2}\left[1+\bz(\bz-\frac{1}{2})\left(1+e^{-2\bz}\right)\right]\mathrm{K}_1(2\bz)\nonumber\\
&&-\left[\frac{\gamma}{2}+\frac{1}{\bz}+\frac{3}{2\bz^2}+\frac{1}{\bz^3}+\frac{1}{2}\ln(4\bz)\right]e^{-2\bz}\nonumber\\
&&+\frac{1}{2\bz^2}e^{-4\bz}+\left(1-\frac{1}{2}e^{2\bz}\right)\mathrm{Ei}(-4\bz).\nonumber
\eea
The above equations make use of the modified Bessel functions of the second kind $\mathrm{K}_n(x)$ and the exponential integral function $\mathrm{Ei}(x)$~\cite{math}. The close form expression Eq.~(\ref{dvLs}) is compared for $s=2$ with the integral form of Eq.~(\ref{self2}) in Fig.~\ref{fig5}.

First of all, we note that the potential given by Eq.~(\ref{dvLsN}) originates from the ion deficiency in the membrane medium $z<0$, and it is known to bring a purely repulsive contribution to the potential $\delta v_0(\bz)$~\cite{PRE}. This repulsive part of the potential marks the upper boundary of the self energy curves in Fig.~\ref{fig5}. The second part $\delta v_{0c}(\bz)$ of Eq.~(\ref{dvLsC}) induced by the surface charge is purely negative. As shown in Fig.~\ref{fig5}, this term brings in turn a net attractive contribution to the self energy. In the asymmetrically distributed salt system considered in the present work, it will be shown that electrostatic correlation effects are mainly driven by the competition between these two opposite mechanisms. This competition will be thoroughly investigated below.

For small separations from the surface $\bz\ll1$, Eq.~(\ref{dvLsC}) takes the asymptotic form
\bea\label{dvLsSz}
\frac{\delta v_0(\bz)}{\Gamma}&\simeq&\frac{1}{3}+\frac{\bz}{8}\left[4\gamma-3+4\ln(\bz)\right]\\
&&-\frac{1}{3s^2}\left\{12\ln(2)-7-\frac{\bz}{4}\left[5+12\gamma+12\ln(\bz)\right]\right\}\nonumber\\
&&+O(\bz^2).\nonumber
\eea
The equation~(\ref{dvLsSz}) shows that the potential $\delta v_0(\bz)$ exhibits a linear decay with the distance $\bz$ close to the interface. We recognize in the first term of this expression the energetic cost $\ell_B\kappa_b/3$ to drive an ion from the bulk to a neutral interface separating a membrane medium free of ions and a salt solution of ionic strength $\kappa_b^2$~\cite{PRE}. Moreover, the contribution from the surface charge is shown to lower this barrier at the interface by the amount $-(12\ln2-7)\Gamma/(3s^2)\simeq-0.44\ell_B/(\kappa_b\mu^2)$, and also to give rise to a potential minimum located between the maximum counterion concentration at $\bz=0$ and the bulk region $\bz=\infty$. By comparing the magnitude of these two terms, one finds that at the particular value $\mu^{-1}\simeq\kappa_b$ where the thickness of the interfacial counterion layer $\mu$ becomes equal to the radius of the ionic cloud $\kappa_b^{-1}$ around a central charge in the bulk area, the potential vanishes on the surface, $\delta v_0(0)=0$. Furthermore, the self energy profile $\delta v_0(\bz)$ remains purely attractive for higher surface charge values or lower bulk ion concentrations. This point is also illustrated in Fig.~\ref{fig5}.

We note that one can also express the competition between the bulk and interfacial solvation effects in terms of the density of the interfacial counterion layer $c\simeq 2\pi\ell_B\sigma_s^2$~\cite{netzcoun} and the bulk salt density as $4\rho_b<c$. This inequality characterizes the regime where the solvation of ions by the interfacial counterion layer attracting them towards the surface overcomes the strength of the interfacial salt screening loss driving the charges far away from the interface. We finally note that the parameter domain where the potential $\delta v_0(\bz)$ is negative can be also rewritten in terms of the balance between the bulk ion concentration and the surface charge as $\rho_b<\pi\ell_B\sigma_s^2/2$. It is interesting to note that this inequality is independent of the ion valency. For instance, for a dilute salt solution with concentration $\rho_b=0.01$ M, this inequality becomes an equality for the characteristic surface charge $\sigma_s=7.5\times10^{-2}$ e $\mathrm{nm}^{-2}$.

In the opposite regime of large separations from the surface $\bz\gg1$, the potential Eq.~(\ref{dvLs}) takes the asymptotic form
\bea\label{dvLsLz}
\frac{\delta v_0(\bz)}{\Gamma}&\simeq&\frac{e^{-2\bz}}{2\bz}\left\{1-s^{-2}\left[\frac{7}{4}+\gamma\bz+\bz\ln(4\bz)\right]\right\}\\
&&+O\left(e^{-4\bz}\right).\nonumber
\eea
In the strict limit $\bz\to\infty$, the leading term of this asymptotic law reads $\delta v_0(\bz)/\Gamma\simeq-s^{-2}\ln(4\bz)e^{-2\bz}$. This shows that far enough from the interface, the attractive surface charge contribution will always dominate the repulsive solvation force associated with the salt screening loss in the proximity of the interface.

In the opposite small $s$ (or the Gouy-Chapman) regime, Eq.~(\ref{self2}) yields
\bea\label{dvgc}
\frac{\delta v_0(\bz)}{\Gamma}&=&\frac{1}{2\bz}\left\{e^{-2\bz}-\left[\gamma+\ln(4\bz)-\mathrm{Ei}(-4\bz)\right]\bz\hspace{1mm}\mathrm{csch}^2(\bz)\right\}\nonumber\\
&&+O(s).
\eea
In Fig.~\ref{fig5}, it is shown that the limiting law Eq.~(\ref{dvgc}) is purely negative. One also notices that this asymptotic limit marks the lower boundary of the self energy curves. For small separations from the surface $\bz\ll1$, the potential decays algebraically with increasing distance,
\be\label{dvgcSz}
\frac{\delta v_0(\bz)}{\Gamma}\simeq-\frac{3}{2\bz}+1-\frac{\bz}{9}+O(\bz^3),
\ee
and for large separations $\bz\gg1$, it is screened exponentially,
\be\label{dvgcLz}
\frac{\delta v_0(\bz)}{\Gamma}\simeq-2\left[\gamma+\ln(4\bz)\right]e^{-2\bz}+O\left(e^{-4\bz}\right).
\ee

The large distance asymptotic limit $\bz\gg1$ of Eq.~(\ref{self2}) can be computed for an arbitrary finite value of $s$ in the form
\bea\label{dvLz}
\frac{\delta v_0(\bz)}{\Gamma}&\simeq&2\gamma_c^2(s)e^{-2\bz}\left[e^{4\bz}\mathrm{Ei}(-4\bz)-\ln(4\bz)-\gamma\right]\\
&&+O\left(e^{-4\bz}\right).\nonumber
\eea
This asymptotic law is reported in Fig.~\ref{fig5} for $s=1$. Noting that $\lim_{x\to\infty}e^{x}\mathrm{Ei}(-x)=-1/x$, the strict large distance limit $\bz\to\infty$ of Eq.~(\ref{dvLz}) is obtained as $\delta v_0(\bz)\simeq-2\Gamma\gamma_c^2(s)e^{-2\bz}\ln(4\bz)$, which is a purely negative function. One can verify that this equality consistently recovers the leading terms of Eq.~(\ref{dvLsLz}) and Eq.~(\ref{dvgcLz}) in the limits of large and small $s$, respectively. The relation~(\ref{dvLz}) confirms the conclusion that we previously reached for weak surface charges : in the presence of an arbitrary finite surface charge and salt concentration, the ionic self-energy will always posses an attractive branch far enough from the interface.
\begin{figure}
(a)\includegraphics[width=1.12\linewidth]{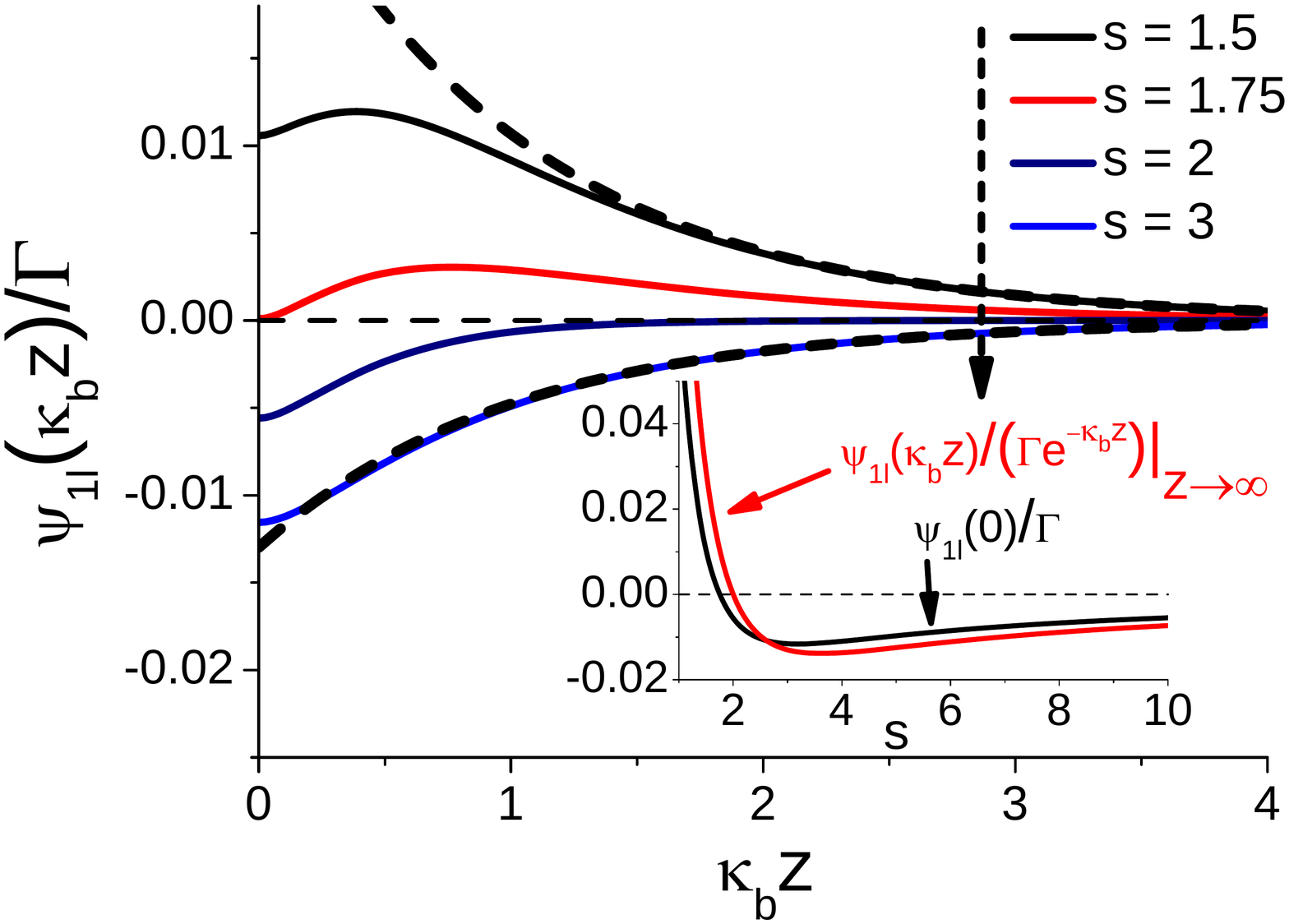}
\caption{(Color online) (a) One-loop correction to the external potential Eq.~(\ref{pot1l}) renormalized by the coupling parameter $\Gamma$ against the dimensioneless distance $\kappa_bz$. The dashed black lines denote the large distance asymptotic limit Eq.~(\ref{psias}). The inset displays the surface potential and the renormalization factor behind Eq.~(\ref{psias}) against the parameter $s$.}
\label{fig6}
\end{figure}

We will now investigate the behaviour of the one-loop correction to the external potential $\psi_{1l}(\bz)$ with respect to the parameter $s$. Evaluating first the integral in Eq.~(\ref{pot1l0}) with Eqs.~(\ref{pmf}), (\ref{eq022}), and~(\ref{self}), one gets for the one-loop correction to the external potential
\be\label{pot1l}
\psi_{1l}(z)=\frac{q^2}{4}\Gamma\mathrm{csch}\left[\bar{z}-\ln\gamma_c(s)\right]\int_1^\infty\frac{\mathrm{d}u}{u^2-1}F(\bz,u),
\ee
with the auxiliary function
\bea\label{F}
F(\bz,u)&=&\frac{2+s^2}{s\sqrt{1+s^2}}-\bd\left(\frac{1}{u}+2u+\frac{2+3s^2}{s\sqrt{1+s^2}}\right)\\
&&+\frac{\bd}{u}e^{-2u\bz}+\left(\bd e^{-2u\bz}-1\right)\coth\left[\bz-\ln\gamma_c(s)\right].\nonumber\\
\eea
We display in the main plot of Fig.~\ref{fig6} the potential profile Eq.~(\ref{pot1l}) for various values of $s$. One sees in this plot that the behavior of the potential is mainly characterized by an interpolation between two regimes where the function $\psi_{1l}(\bz)$ changes its sign. In the first regime of large $s$ (or weak surface charges and high salt concentrations) characterized by  a positive energy barrier $\delta v_0(\bz)$ (see Fig.~\ref{fig5}), the resulting interfacial ion depletion responsible for a local ionic screening deficiency of the external potential increases the amplitude of the negative mean-field potential $\varphi(\bz)$. This regime is the reminiscent of the behavior observed in the previous part for the electrolyte system in contact with a dielectrically discontinuous wall (see Fig.~\ref{fig4}(a)). Loosely speaking, in this regime, the MF theory overestimates the screening of the external potential. In the second regime of strong surface charges or dilute electrolytes corresponding to small values of $s$ and an attractive self energy (see again Fig.~\ref{fig5}), the compact counterion layer formation at the interface is associated with a local ionic screening excess with respect to the bulk solution. This means that the MF theory underestimates the ionic screening in this parameter range, which is corrected with a positive $\psi_{1l}(\bz)$. The interpolation between both regimes will be quantitatively studied below.

Expanding the function $F(\bz,u)$ in Eq.~(\ref{F}) in inverse powers of $s$ and carrying out the integral in Eq.~(\ref{pot1l}) at $\bz=0$, the surface potential follows in the form of a series as
\bea\label{pot1lS}
\psi_{1l}(0)&=&\frac{q^2}{4}\Gamma\left\{-0.227s^{-1}+0.856s^{-3}-0.645s^{-5}\right.\\
&&\hspace{0.9cm}\left.+0.539s^{-7}-0.473s^{-9}+O\left(s^{-11}\right)\right\}.\nonumber
\eea
From the relation~(\ref{pot1lS}), one finds that the one-loop correction to the surface potential vanishes at the value $s\simeq1.75$ and remains negative for higher values of $s$. This is illustrated in the inset and the main plot of Fig.\ref{fig6}. The function $\psi_{1l}(0)$ also exhibits a minimum located at $\simeq3.19$ before decaying to zero with increasing $s$.
\begin{figure}
a)\includegraphics[width=.95\linewidth]{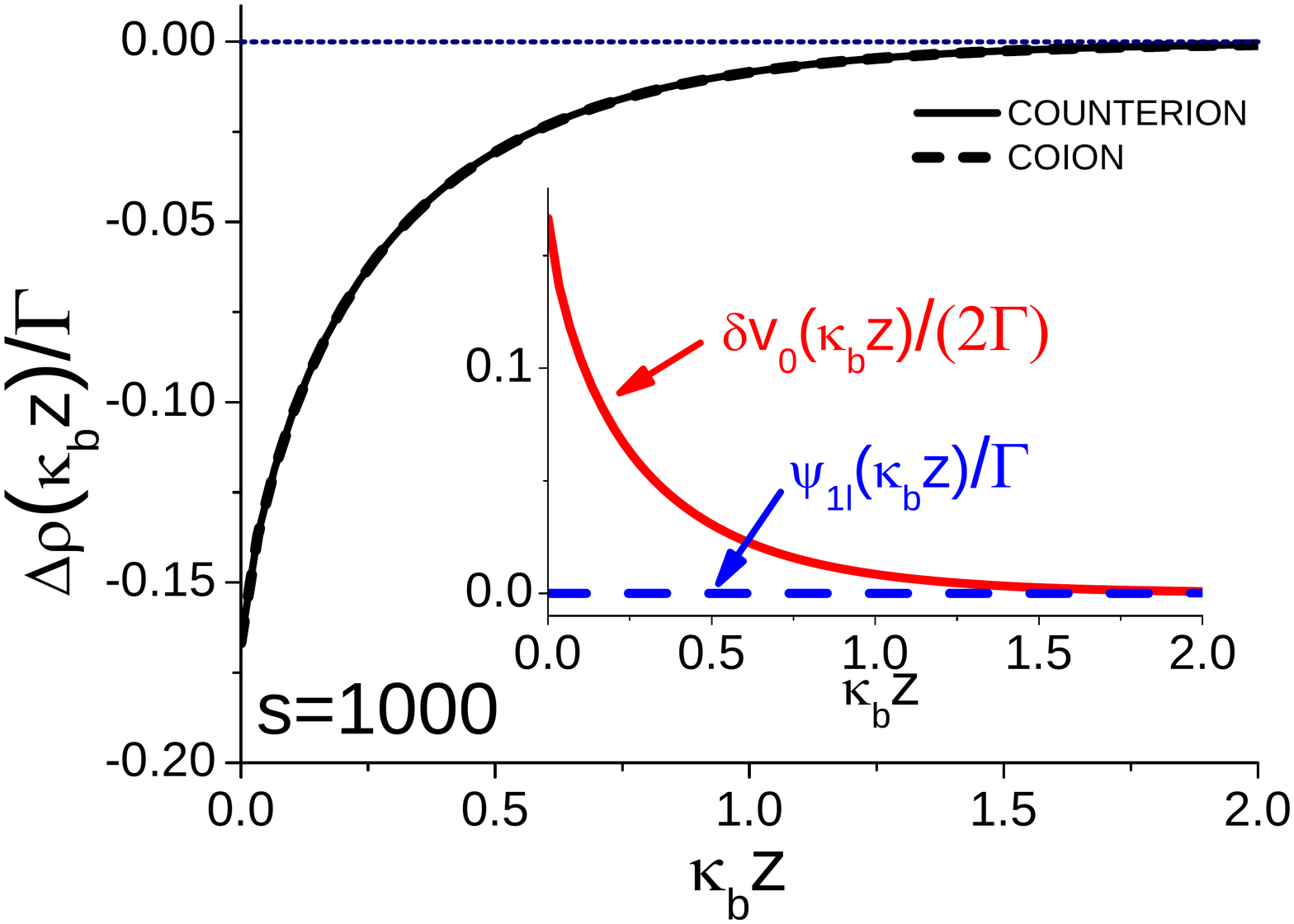}
b)\includegraphics[width=0.95\linewidth]{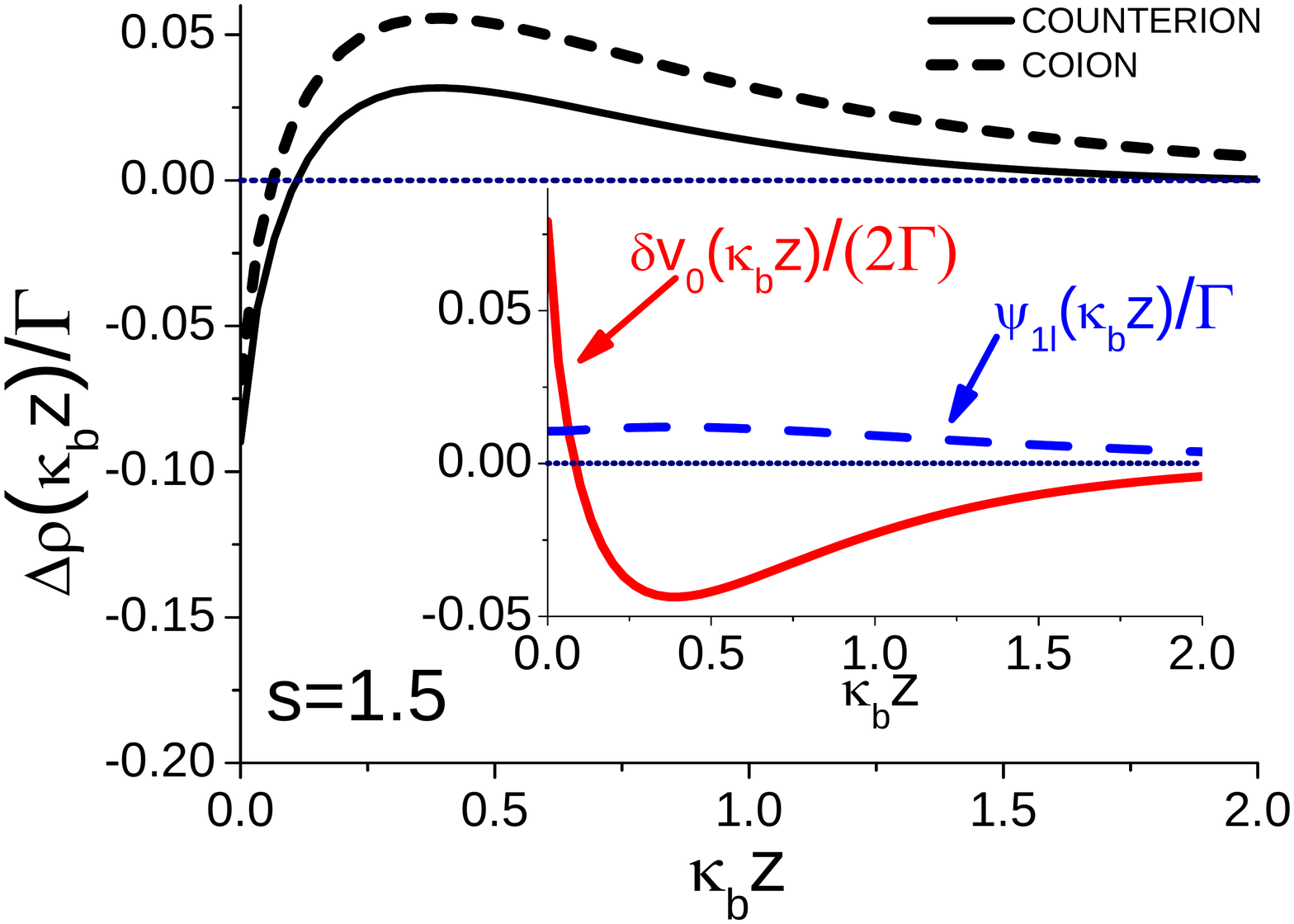}
c)\includegraphics[width=.95\linewidth]{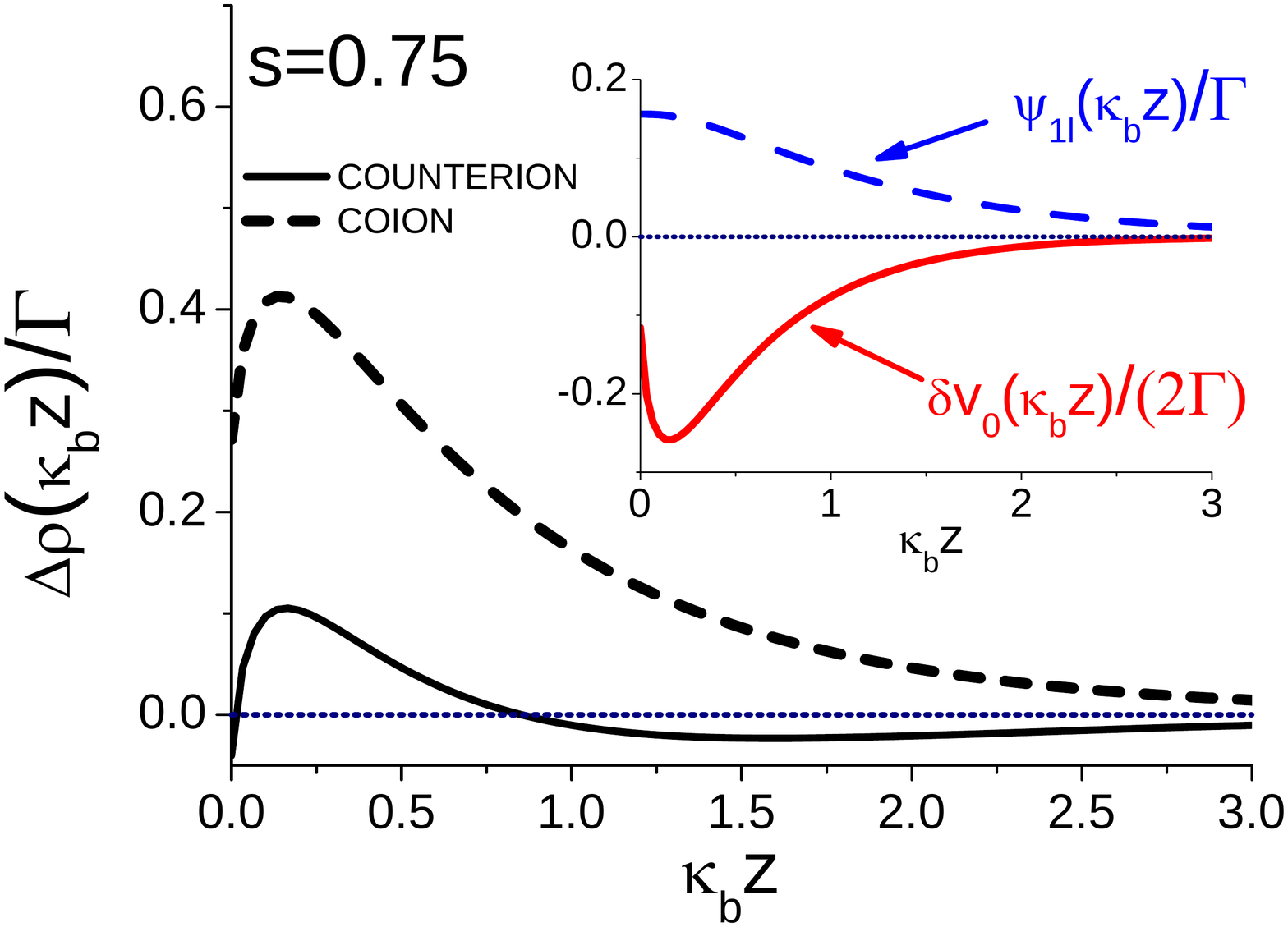}
\caption{(Color online) One-loop correction to counterion (solid lines) and coion densities (dashed lines) from Eq.~(\ref{den1loopcor}) for (a) $s=1000$, (b) $s=1.5$, and (c) $s=0.75$. The inset displays the ionic self energy and the one-loop correction to the external potential for the same parameters.}
\label{fig7}
\end{figure}

For large distances from the charged interface $\bz\gg1$, the one-loop correction to the external potential Eq.~(\ref{pot1l}) takes the simple asymptotic form
\be\label{psias}
\psi_{1l}(\bz)\simeq\frac{q^2}{2}\Gamma\gamma_c(s)\mathrm{I}(s)e^{-\bz},
\ee
where we introduced the auxiliary function
\bea
\mathrm{I}(s)&=&\int_1^\infty\frac{\mathrm{du}}{u^2-1}\left\{\frac{2+s^2}{s\sqrt{1+s^2}}-1\right.\\
&&\hspace{2cm}\left.-\bd\left(\frac{1}{u}+2u+\frac{2+3s^2}{s\sqrt{1+s^2}}\right)\right\}.\nonumber
\eea
The asymptotic form in Eq.~(\ref{psias}) is displayed in Fig.~\ref{fig6} by black dashed lines. The function behind the exponential in this equation can be evaluated for large values of $s$ as
\bea\label{IS}
\gamma_c(s)\mathrm{I}(s)&=&-0.153s^{-1}+0.743s^{-3}-0.628s^{-5}\nonumber\\
&&+0.542s^{-7}+O\left(s^{-9}\right).
\eea
One finds that this function vanishes at the particular value $s\simeq2$, that is, the one-loop correction to the external potential changes its sign and becomes overall positive at $\mu=2\kappa_b^{-1}$, before reaching a minimum located at $s\simeq3.62$. We note that interestingly, this sign reversal takes place at the characteristic surface charge $\sigma_s=\sqrt{\rho_b/(2\pi\ell_B)}$, which is twice lower than the surface charge where the self energy $\delta v_0(z)$ becomes totally attractive.

The deviations of the one-loop density from the MF density $\rho_\pm^{MF}(\bz)=\rho_be^{-V_w(\bz)\mp\varphi(\bz)}$ can be expressed as
\be\label{den1loopcor}
\Delta\rho_\pm(\bz)\equiv\frac{\rho_\pm(\bz)-\rho_\pm^{MF}(\bz)}{\rho_\pm^{MF}(\bz)}=-\frac{q^2}{2}\delta v_0(\bz)\mp\psi_{1l}(\bz).
\ee
We illustrate in the main plots of Fig.~\ref{fig7} the one-loop correction to ion densities renormalized by the coupling parameter $\Gamma$ for $s=0.75$, $1.5$, and $1000$. The one-loop corrections $\delta v_0(\bz)$ and $\psi_{1l}(\bz)$ are also shown in the inset of the same figures. In the case $s=1000$ corresponding to weak surface charges where the amplitude of the potential $\psi_{1l}(\bz)$ remains vanishingly small with respect to the repulsive self energy $\delta v_0(\bz)$ (see the inset of Fig.~\ref{fig7}(a)), the repulsive solvation force induced by the interfacial salt screening loss is the only effect in play. Consequently, the one-loop correction lowers the MF density of both coions and counterions.

In the case $s=1.5$ of Fig.~\ref{fig7}(b) corresponding to a stronger surface charge or lower salt density where the interfacial counterion layer becomes dense enough to give rise to a strongly attractive branch of the self energy potential $\delta v_0(\bz)$, correlation effects increase the MF density of both types of ions. In this range of the parameter $s$ where the potential $\psi_{1l}(\bz)$ has a positive and a considerably large amplitude, the main plot of Fig.~\ref{fig7}(b) shows that interestingly, the additional attraction induced by the ionic self energy is amplified by the external potential correction for coions, whereas the same attraction is partially cancelled for counterions. We show in Fig.~\ref{fig7}(c) that this effect is even stronger in the range $s\lesssim1$ where the amplitude of the potential $\psi_{1l}(\bz)$ becomes comparable with the amplitude of the self energy $\delta v_0(\bz)$. This indicates that correlations attenuate in this parameter regime charge separation in the interfacial region.

We also see in Fig.~\ref{fig7}(c) that for $s\lesssim1$, the contribution from the one-loop correction to the external potential leads to a reduction of the MF counterion density at large distances from the interface. This peculiarity can be easily understood by comparing Eq.~(\ref{psias}) with the large $\bz$ asymptotic limit of the ionic self energy Eq.~(\ref{dvLz}). One sees that the one-loop correction to the external potential $\psi_{1l}(\bz)$ is longer ranged than the self energy potential $\delta v_0(\bz)$. Thus, in the asymptotic limit $\bz\gg1$, the former brings the main contribution to the one-loop corrections for the MF density in Fig.~\ref{fig7}(c).

\begin{figure}
\includegraphics[width=1.15\linewidth]{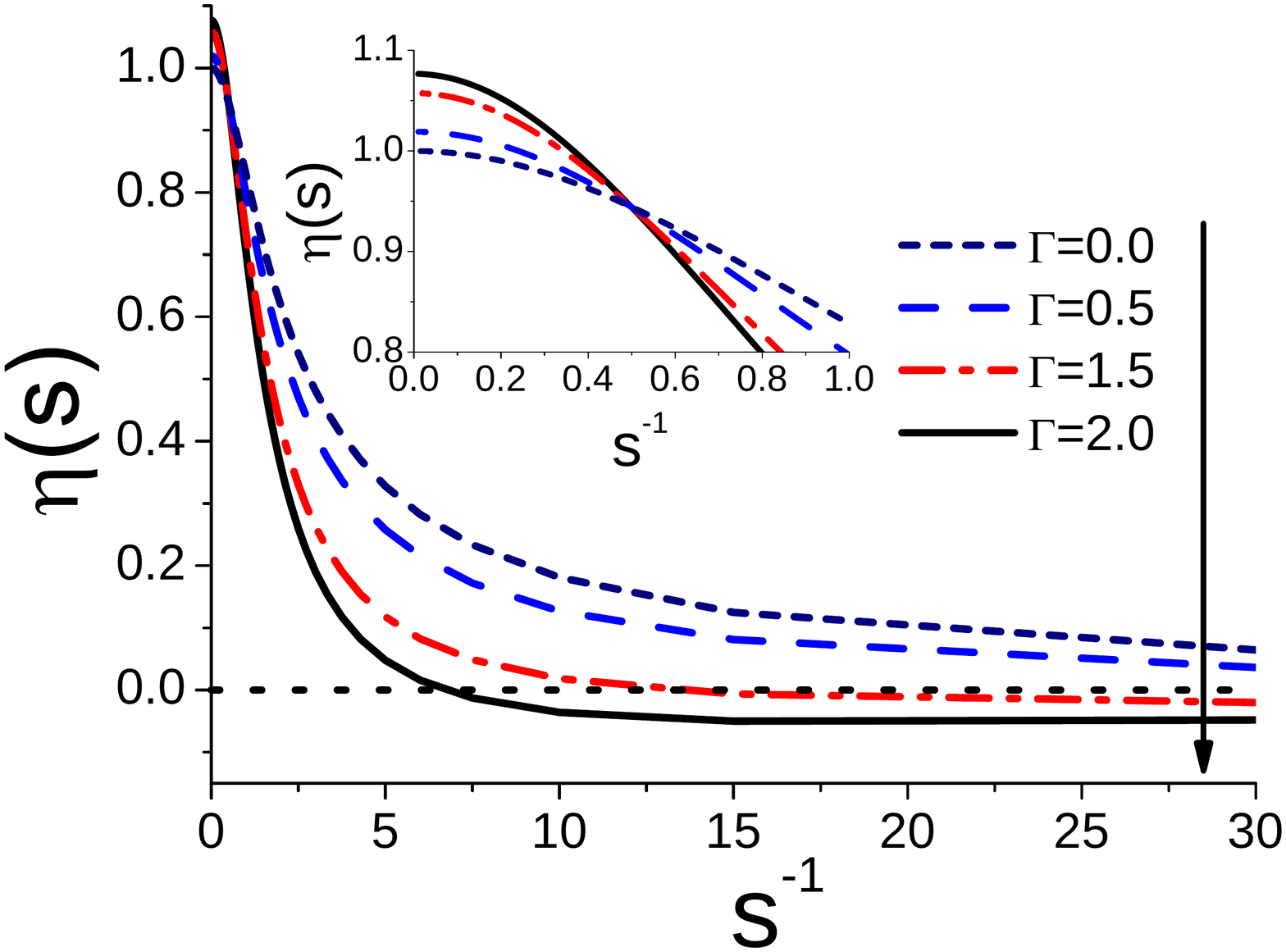}
\caption{(Color online) Charge renormalization factor against $s^{-1}$ for several values of the coupling parameter $\Gamma$.}
\label{fig8}
\end{figure}

Taking now into account the large distance asymptotic limit $\bz\gg1$ of the MF potential Eq.~(\ref{pmf}) that can be written as $\varphi(\bz)\simeq-4\gamma_c(s)e^{-\bz}$, the total one-loop external potential $\phi_{1l}(\bz)\equiv\varphi(\bz)+\psi_{1l}(\bz)$ reads for $\bz\gg1$
\be\label{phi1lLz}
\phi_{1l}(\bz)\simeq-\frac{2}{s}\eta(s)e^{-\bz},
\ee
where we introduced a charge renormalization factor dressed by electrostatic correlation in the form
\be\label{eta}
\eta(s)=2s\gamma_c(s)\left[1-\frac{q^2\Gamma}{8}\mathrm{I}(s)\right].
\ee
We illustrate in Fig.~\ref{fig8} the charge renormalization factor Eq.~(\ref{eta}) against the parameter $s^{-1}\propto\sigma_s$ for different values of the coupling parameter $\Gamma$. The first point to be noted in this plot is the intersection between the curves for different values of $\Gamma$ at $s=2$ (see the inset). This point corresponds to the parameter range where the asymptotic large $\bz$ limit of the one-loop correction to the external potential vanishes and all curves collapse onto the MF charge renormalization factor. One also sees that for $s>2$ (the left portion of the intersection point), $\eta(s)$ increases with $\Gamma$. This behavior can be easily understood by noting that this parameter regime was shown above to correspond to a weak interfacial ionic screening deficiency resulting in a negative one-loop correction to the negative mean-field potential. Furthermore, by taking the vanishing surface charge limit of Eq.~(\ref{eta}), one finds $\lim_{s\to\infty}\eta(s)=1+3.8\times10^{-2}\hspace{0.5mm}q^2\Gamma$. In other words, electrostatic correlations interestingly yield a finite correction to the charge renormalization factor even in the limit of a vanishing fixed charge distribution.

In the second regime $s<2$ corresponding to stronger surface charges, it is seen that $\eta(s)$ changes its trend and starts to decrease with increasing coupling parameter $\Gamma$. Indeed, it was shown above that this regime is characterized by a compact counterion layer associated with an interfacial screening excess and a positive one-loop correction to the negative MF potential. This aspect explains the trend of the charge renormalization factor for $s<2$.

Furthermore, one notices that at a particular value of $s$, the factor $\eta(s)$ changes its sign and becomes negative, resulting in a sign reversal of the total one-loop potential in Eq.~(\ref{phi1lLz}), which becomes positive. The reversal of the sign of the external potential above a characteristic surface charge is a signature of the \textit{charge inversion} phenomenon. We note that this effect was also observed in Ref.~\cite{1loop} for the symmetrically distributed electrolyte system. However, we also emphasize that the coupling parameter range where this effect takes place in Fig.~\ref{fig8} is beyond the validity of the one-loop approximation. Indeed, we verified that in this electrostatic coupling regime, even the perturbative solution scheme of the SC equations~(\ref{eq5})-(\ref{eq6}) introduced in section~\ref{anan} does not converge. Furthermore, the MF counterion densities reach in this charge density regime unrealistic values beyond the close packing, an artefact known to originate from the absence of ionic excluded volume effects in the model Hamiltonian of Eq.~(\ref{HamFunc})~\cite{Boruk}. The inclusion of hard-core effects known to reduce the interfacial counterion densities is also expected to shift the charge reversal point to larger surface charges. Consequently, it is clear to us that in Fig.~\ref{fig8}, the curves with $\Gamma>1$  have no quantitative reliability for large values of $s^{-1}$. That being said, we believe that this result still presents some qualitative interest since the effect observed in Fig.~\ref{fig8} might be the precursor of the actual charge inversion phenomenon in asymmetrically distributed salt systems. This point needs however to be verified by comparisons with MC simulations.

\section{Summary and Conclusion}

In this article, we investigated electrostatic correlation effects in symmetric electrolytes in contact with charged planar interfaces separating the solvent region from a membrane area free of ions. To this aim, we introduced in the first part of the work two computational approaches to solve the electrostatic SC equations derived in Ref.~\cite{netzvar} in the presence of dielectric discontinuities where the one-loop theory fails. Then, we compared in the second part the theoretical ion density profiles obtained from these computation schemes with the results of our MC simulation data in order to determine the validity domain of the SC equations at neutral dielectric interfaces. It was shown that the DH theory that neglects the interfacial variations of the ionic screening exhibits a quantitative accuracy up to the characteristic bulk density $\rho_b\simeq 0.01$ M, while the SC theory remains accurate up to $\rho_b\simeq 0.2$ M, thus improving the quantitative accuracy of the DH theory by one order of magnitude in ionic strength. The deviations of the SC results from MC simulation data at $\rho_b\gtrsim 0.2$ M may be either due to electrostatic correlation effects, or hard-core effects that become relevant in this concentration regime. This point can be enlightened in a future work by solving the extended SC equations of Ref.~\cite{jstat} that account for the excluded volume of ions.

Within the validity regime of SC equations, we also validated the accuracy of a restricted variational scheme introduced in~\cite{PRE} for slit nanopore systems. This observation may explain the success of similar methods frequently used in nanofiltration studies in order to predict experimental salt rejection rates~\cite{yarosch,YarII,Szymczyk}. Furthermore, we computed within the WKB formalism the interaction energy between a charged rigid polymer and a neutral dielectric interface. We showed that due to the interfacial ionic screening deficiency neglected in the DH approximation, the energetic cost to bring a polymer from the bulk to the proximity of the interface is characterized by a different scaling law with the distance from the interface as well as a significantly higher amplitude than the prediction of the DH theory~\cite{netzdh}. This observation relevant for protein-surface interactions could be verified with MC simulations.

In the third part of the article, we considered the case of ions at charged interfaces. For ions in the proximity of a weakly charged dielectric wall, we showed that the main correlation effect is a strong interfacial salt exclusion resulting in an ionic screening deficiency that increases the amplitude of the negative MF external potential. The latter effect strengthens the counterion attraction and coion repulsion far away from the interface. Thus, correlation effects in weakly charged membrane nanopores are expected to amplify the charge separation phenomenon induced by Donnan exclusion. Comparing the ion density and external potential profiles obtained from the SC scheme with MC simulation data of Ref.~\cite{torrie}, we also showed that the SC theory is able to accurately handle these correlation effects in a parameter regime where the MF theory exhibits a significant deviation from the MC results.

The forth part of the work dealt with electrostatic correlation effects in dielectrically homogeneous systems at the one-loop level. We first expanded the SC theory at one-loop order and found that one-loop corrections in this system are driven by the competition between the interfacial salt screening loss driving the ions towards the bulk area, and the counterion screening excess induced by the interfacial counterion layer attracting them to the interface. This competition can be quantified in terms of the balance between the characteristic thickness of the interfacial counterion layer $\mu$ and the radius of the ionic cloud $\kappa_b^{-1}$ around a central charge in the bulk region. Namely, in the presence of a weak surface charge corresponding to a diffuse counterion layer $\mu\gg\kappa_b^{-1}$, the surface-repulsive salt screening effect is the dominant mechanism. In this case, correlation effects result in a net interfacial exclusion of both coions and counterions, and a weak negative correction to the MF potential.

Decreasing the bulk salt concentration or equivalently increasing the surface charge to the range $\sigma_s\simeq\sqrt{2\rho_b/(\pi\ell_B)}$, the thickness of the counterion layer $\mu$ becomes comparable with the ionic cloud radius $\kappa_b^{-1}$. As a result, the surface-attractive counterion screening effect starts to set on and correlation effects increase in this regime the MF density of both types of ions. Moreover, due to the strong counterion excess close to the surface, the one-loop external potential correction also acquires in this surface charge regime a positive and large amplitude. Consequently, while a further increase of the surface charge will result in a amplification of the MF coion density in the whole half space $z>0$ and the MF counterion density close to the surface, the latter will be attenuated by correlation effects outside the interfacial region.

It was also shown that if one reaches a high enough surface charge value, the one-loop theory predicts charge inversion for $\Gamma\gtrsim1$. However, as we stressed in the main text, this result should be considered with caution. Indeed, this parameter regime stays well beyond the validity range of the one-loop theory. Hence, comparison with MC simulations is needed in order to verify whether this observation might be the precursor of the actual charge inversion effect in asymmetrically distributed salt systems.

The one-loop theory of asymmetrically partitioned electrolytes presented in the final part of this work bridges a gap between the DH theory of symmetric salts at neutral interfaces~\cite{netzdh} and the one-loop theory of counterions in contact with a charged surface~\cite{netzcoun}. Furthermore, we note that the theoretical concepts introduced in the present work can be applied to more complicated geometries such as cylindrical ion channels~\cite{PRL}, or the charged Yukawa model studied in Ref.~\cite{jstat} in order to evaluate the importance of the ionic excluded volume on the correlation effects that we have investigated. We would also like to establish in a future work the validity regime of the SC equations~(\ref{eq5}) and~(\ref{eq6}) with respect to the surface charge by running extensive MC simulations of ions in contact with charged interfaces.
Finally, we note that the perturbative SC scheme can also account for electrostatic correlation effects in more complicated electrostatic systems where the one-loop theory fails, such as polymer brushes at charged dielectric interfaces or dielectric polyelectrolytes in ionic solutions. We would like to treat these cases in future works.
\\
\acknowledgements  Sahin Buyukdagli thanks John Palmeri for having mentioned the possibility to solve the SC equations within the WKB method, which incited the author to scan the literature and find the Ref.~\cite{japan} where the idea was originally introduced. This work has been in part supported by The Academy of Finland through its COMP CoE and NanoFluid grants.
\smallskip
\appendix
\section{Perturbative solution of the closure equations within WKB approach}
\label{persol}

A perturbative solution to the closure equations~(\ref{saltSC}) and~(\ref{kerself}) was derived in Ref.~\cite{japan} for simple interfaces in the case $\e_m=0$. We will rederive in this Appendix this result obtained for finite $\e_m$. The first step consists in writing the screening function of Eq.~(\ref{saltSC}) in the form
\be\label{itkap}
\kappa_0^2(z)=\kappa_b^2\left[1+\delta n_0(z)\right],
\ee
where $\delta n_0(z)$ is a "small" perturbation of the screening length induced by the image potential that we choose as the screened image potential within the \textit{undistorted ionic atmosphere approximation}~\cite{yarosch},
\be
\delta n_0(z)=\mathrm{exp}\left(-\frac{q^2\ell_B\Delta_0}{4z}e^{-2\kappa_bz}\right)-1.
\ee
We note that this choice is motivated by the very weak dependence of the image potential profile on the input image potential, a good convergence property of the iterative computation scheme observed in the numerical solutions of the closure equations~(\ref{saltSC}) and~(\ref{kerself}). Furthermore, the naught in Eq.~(\ref{itkap}) denotes the input function at the zeroth order iterative level. By injecting this function into the electrostatic potential~(\ref{eq10}) and expanding in $\delta n_0(z)$, one gets after some algebra
\be\label{potit1}
\delta v(z)=\frac{\ell_B\Delta_0}{2z}e^{-2\kappa_bz}+\delta v_1(z),
\ee
with the correction to the WC image potential
\bea
\delta v_1(z)&=&\ell_B\kappa_b\left\{1-\mathrm{exp}\left(-\frac{q^2\ell_B\Delta_0}{8z}e^{-2\kappa_bz}\right)\right\}\\
&&-\frac{\ell_B}{2}\kappa_b\Delta_0\delta n_0(z)e^{-2\kappa_bz}\nonumber\\
&&+\ell_B\Delta_0\kappa_b^2\Gamma(0,2\kappa_bz)\int_0^z\mathrm{d}z'z'\frac{d\delta n_0(z')}{dz'},\nonumber
\eea
where $\Gamma(a,x)$ is the incomplete Gamma function~\cite{math}. By injecting now Eq.~(\ref{potit1}) into Eq.~(\ref{den}), the ion density profile at the first iterative level finally takes the form
\be\label{den1}
\frac{\rho_1(z)}{\rho_b}=\mathrm{exp}\left(-\frac{q^2\ell_B\Delta_0}{4z}e^{-2\kappa_bz}\right)\left[1-\frac{q^2}{2}\delta v_1(z)\right].
\ee
We note that in the limit $\e_m=0$, the expression~(\ref{den1}) reduces to the one derived in Ref.~\cite{japan}.
\section{Third order perturbative solution of SC equations for neutral interfaces}
\label{3th}

We will give in this Appendix the third order perturbative correction for the solution of the SC equation~(\ref{eq6}) for a vanishing surface charge $\sigma(z)=0$. Because this case corresponds to a zero external potential, the components of the correction to the Green's function $\delta v_{nm}$ with $m>0$ vanish. As in the computation of the second order calculation presented in section~\ref{anan}, the analytical task consists in injecting into Eq.~(\ref{eq332}) the expansion of the Green's function~(\ref{eq337}) and keeping only the terms up to $\lambda_v^3$. One obtains, in addition to the first and second order corrective terms in Eqs.~(\ref{eq339}) and~(\ref{eq341}), the third order correction in the form
\bea
\delta v_{30}(z)&=&-\frac{q^4}{4}\rho_b\int\mathrm{d}z_1n_0(z_1)\left[q^2\delta v_{10}^2(z_1)-4\delta v_{20}(z_1)\right]\nonumber\\
&&\hspace{2cm}\times I_2(\br,\br',z_1)e^{-V_w(z_1)}\nonumber\\
&&-2q^6\rho_b^2\int\mathrm{d}z_1\mathrm{d}z_2n_0(z_1)\delta v_{10}(z_1)\delta n_0(z_2)\nonumber\\
&&\hspace{1.5cm}\times \left[I_3(\br,\br',z_1,z_2)+I_3(\br',\br,z_1,z_2)\right]\nonumber\\
&&\hspace{1.5cm}\times e^{-V_w(z_2)}]\nonumber\\
&&-8q^6\rho_b^3\int\mathrm{d}z_1\mathrm{d}z_2\mathrm{d}z_3\delta n_0(z_1)\delta n_0(z_2)\delta n_0(z_3)\nonumber\\
&&\hspace{3cm}\times I_4(\br,\br',z_1,z_2,z_3),
\eea
where we defined the new function
\bea
&&I_4(\br,\br',z_1,z_2,z_3)=\int\frac{\mathrm{d}^2\bk}{4\pi^2}e^{i\bk\cdot\left(\br_\parallel-\br'_\parallel\right)}\tv_0(z,z_1,k)\nonumber\\
&&\hspace{4cm}\times\tv_0(z_1,z_2,k)\tv_0(z_2,z_3,k)\nonumber\\
&&\hspace{4cm}\times\tv_0(z_3,z',k).
\eea
Finally, at the third order perturbative level, the ion density Eq.~(\ref{den}) reads
\bea\label{denper3}
\rho(z)/\rho^{(0)}(z)&=&1-\lambda_v\frac{q^2}{2}\delta v_{10}(z)\\
&&+\lambda_v^2\frac{q^2}{8}\left[\delta v_{10}^2(z)-4\delta v_{20}(z)\right]\nonumber\\
&&-\lambda_v^3\frac{q^2}{48}\left[q^4\delta v_{10}^3(z)-12q^2\delta v_{10}(z)\delta v_{20}(z)\right.\nonumber\\
&&\left.\hspace{1.2cm}+24\delta v_{30}(z)\right].\nonumber
\eea
\section{Monte Carlo simulations}
\label{mcdetails}

We present in this Appendix the details of the canonical MC simulations.  During the simulations, the total particle number was fixed at $N_p=4096$, with the corresponding particle density $\bar \rho_{\textrm{MC}}=N_p/(l_xl_yl_z)=\rho_b$, the size of the simulation box $l_x=l_y=(N_p/1.5\rho_b)^{1/3}$ and $l_z=1.5l_x$, and impenetrable walls located at $z=0$ and $z=l_z$. Periodic boundary conditions in the $x$ and $y$ directions were used. The value of $l_z$ that we considered was large enough so that the box geometry was equivalent to the actual single interface geometry in consideration.

The MC simulations were run on nvidia  graphical processing units (GPU) using CUDA algorithm~\cite{cuda}. In the simulations, all the relevant variables were set to reside in GPU RAM memory. By dividing the work in small independent parts that are all executed in parallel, the GPU can provide up to two orders of magnitude improvement in performance. In our system, we have divided the calculation of the energy difference between the present and the trial configuration in small tasks. For each trial move, $2N_p$ threads were used for calculating the individual particle-particle interactions, followed by a sum-reduction to obtain the total energy difference between the present and the new configuration. Finally, only $1$ thread was checking the acceptance conditions and making the necessary updates. Even if some parts of the MC algorithm were not able to fully utilize the GPU, we were able to obtain a speed-up larger than $100$ times.

\end{document}